\documentclass[12pt,a4paper]{article}
\usepackage{setspace}
\usepackage[title,titletoc]{appendix}
\usepackage{graphicx}
\usepackage{graphics,epsfig}
\usepackage{amsfonts,amssymb,amsmath,mathenv}
\usepackage{enumerate}
\usepackage{epstopdf}
\allowdisplaybreaks[3]

\newcommand{\nc}{\newcommand}
\def\foot{\footnote}
\def \bi{\bibitem}
\def \ci{\cite}
\nc{\sr}{\sqrt}
\nc{\fr}{\frac}
\nc{\ov}{\over}
\nc{\x}{\times}
\nc{\cosec}{\textrm{\,cosec\,}} 
\nc{\sech}{\textrm{\,sech\,}}
\nc{\cosech}{\textrm{\,cosech\,}}
\nc{\del}{\partial}
\nc{\dpl}{\partial_+}
\nc{\dm}{\partial_-}
\nc{\dpdp}{\partial_+\partial_+}
\nc{\dmdm}{\partial_-\partial_-}
\nc{\dpdm}{\partial_+\partial_-}
\nc{\dpm}{\partial_\pm}
\nc{\ra}{\rightarrow}
\nc{\lra}{\leftrightarrow}
\nc{\Ra}{\Rightarrow}
\nc{\til}{\tilde}
\nc{\R}{\mathbb{R}}
\nc{\Z}{\mathbb{Z}}
\nc{\C}{\mathbb{C}}

\nc{\al}{\alpha}
\nc{\bet}{\beta}
\nc{\ga}{\gamma}
\nc{\de}{\delta}
\nc{\ep}{\epsilon}
\nc{\ch}{\chi}
\nc{\h}{\eta}
\nc{\thet}{\theta}
\nc{\ka}{\kappa}
\nc{\lam}{\lambda}
\nc{\si}{\sigma}
\nc{\Si}{\Sigma}
\nc{\ta}{\tau}
\nc{\ze}{\zeta}
\nc{\y}{\psi}
\nc{\om}{\omega}
\nc{\vt}{\vartheta}
\nc{\vp}{\varphi}
\nc{\be}{\begin{equation}}
\nc{\ee}{\end{equation}}
\nc{\bd}{\begin{displaymath}}
\nc{\ed}{\end{displaymath}}
\nc{\ba}{\begin{array}}
\nc{\ea}{\end{array}}
\nc{\bs}{\begin{split}}
\nc{\es}{\end{split}}
\nc{\nn}{\nonumber}
\nc{\la}{\label}
\nc{\eref}[1]{{\left(\ref{#1}\right)}}
\nc{\nin}{\noindent}
\nc{\hs}{\hspace}
\nc{\vs}{\vspace}
\nc{\mc}{\mathcal}
\nc{\mf}{\mathfrak}
\nc{\mbb}{\mathbb}
\nc{\trm}{\textrm}
\nc{\tbf}{\textbf}
\nc{\mbf}{\mathbf}
\def \STr {{\trm{Tr}}}
\def \by {\times } 
\nc{\Lag}{\mathcal{L}}
\nc{\Ham}{\mathcal{H}}
\nc{\psu}{\mf{psu}\left(2,2\,|4\right)}
\nc{\Psu}{PSU\left(2,2|4\right)}
\nc{\gi}{g^{-1}}
\nc{\finv}{f^{-1}}
\nc{\YL}{\Psi_{_L}}
\nc{\YR}{\Psi_{_R}}
\nc{\YLO}{\Psi_{_L0}}
\nc{\YRO}{\Psi_{_R0}}
\nc{\hm}{\h^\parallel}
\nc{\hh}{\h^\perp}
\nc{\kasym}{$\ka$-symmetry }
\nc{\com}[2]{[ #1,\,#2 ]}
\nc{\acom}[2]{\{#1 ,\,#2\}}
\nc{\deta}[2]{\mf{L}^{#1}_\h\left(#2\right)}
\nc{\dM}[2]{\mf{L}^{#1}_\M\left(#2\right)}
\nc{\detaop}{\mf{L}_\h}
\nc{\dMop}{\mf{L}_\M}
\nc{\hta}{\hat{\tau}}
\nc{\bra}[1]{\left<#1\right|}
\nc{\ket}[1]{\left|#1\right>}
\nc{\tp}{\tilde{p}}
\nc{\adot}{\dot{a}}
\nc{\bdot}{\dot{b}}
\nc{\cd}{\dot{c}}
\nc{\dd}{\dot{d}}
\nc{\ald}{\dot{\al}}
\nc{\betd}{\dot{\bet}}
\nc{\gad}{\dot{\ga}}
\nc{\ded}{\dot{\de}}
\nc{\bars}{\bar{\si}}
\nc{\M}{X}
\nc{\psl}{\psi_{_L}}
\nc{\psr}{\psi_{_R}}
\nc{\zel}{\zeta_{_L}}
\nc{\zer}{\zeta_{_R}}
\nc{\chl}{\chi_{_L}}
\nc{\chr}{\chi_{_R}}

\def \const {{\rm const}}
\def \la {\label}
\def \bea{\begin{eqnarray}}
\def \eea{\end{eqnarray}}

\def \Str {\trm{STr}}
\def \no {\nonumber}
\def \ov {\over}
\def \tr {{\rm Tr}}
\newcommand{\rf}[1]{(\ref{#1})}
\def \g {\gamma}
\def \tm {\mbb{T}}
\def \ha {\fr{ 1}{ 2}}

\def \vp {\varphi}

\def \H {{\rm H}} 
\def \Tr {{\rm Tr}}
\def \adss  {$AdS_5 \x S^5$\ }
\def \id {\tbf{1}}
\def \su {\mf{su}}
\def \so {\mf{so}}
\def \ep {\epsilon}
\def \bp {\begin{pmatrix}} 
 \def \emp {\end{pmatrix}}
 
\begin{document}

\overfullrule=0pt
\parskip=2pt
\parindent=12pt
\headheight=0in \headsep=0in \topmargin=0in \oddsidemargin=0in

\vspace{ -3cm}
\thispagestyle{empty}
\vspace{-1cm}

\rightline{Imperial-TP-AT-2009-06}


\begin{center}
\vspace{1cm}
{\Large\bf Tree-level S-matrix of  Pohlmeyer reduced form\\
\vspace{.3cm}
 of $AdS_5 \x S^5$  superstring theory
\vspace{1cm}}
\vspace{.2cm}

{B. Hoare\footnote{benjamin.hoare08@imperial.ac.uk }  and A.A. Tseytlin\footnote{Also at
 Lebedev  
Institute, Moscow. tseytlin@imperial.ac.uk }}
\\
\vskip 0.6cm
{\em 
 Theoretical Physics Group \\
 Blackett Laboratory, Imperial College\\
London SW7 2AZ, U.K. }
\end{center}

\setcounter{footnote}{0}
\begin{abstract} \noindent\begin{footnotesize}
With a motivation to find a 2-d Lorentz-invariant solution 
of the   \adss  superstring  we continue the study 
of the Pohlmeyer-reduced form of this theory.
The reduced theory is constructed from    currents
 of the superstring  sigma model 
and is classically  equivalent to it.
Its action is that of 
  $G/H=Sp(2,2)\x Sp(4)/[SU(2)]^4$  gauged  WZW  model 
 deformed by  an integrable potential   and coupled to fermions.
This theory is UV finite and  is conjectured  to be 
 related to the superstring theory also at  the quantum level. 
Expanded near the trivial vacuum it has  the same elementary 
excitations  
(8+8 massive bosonic and fermionic  2-d degrees of freedom) as the 
\adss  superstring in the  $S^5$  light-cone gauge or near plane-wave 
expansion. In contrast to  the superstring case,  
the  interaction  terms in the reduced  action 
are manifestly 2-d Lorentz invariant. 
 Since the  theory is integrable, 
its  S-matrix  should be effectively determined by the two-particle 
scattering. Here we explicitly 
compute the   tree-level two-particle 
S-matrix for the elementary excitations 
of the reduced theory. We find that this S-matrix 
  has the same 
 index  structure and  group factorization properties 
as the superstring S-matrix  computed in  hep-th/0611169 
but has   simpler coefficients,  depending  only on the difference of two 
rapidities. 
While the gauge-fixed form of  the reduced   action  has only 
the bosonic 
 $[SU(2)]^4$ part of the $PSU(2|2) \times PSU(2|2)$ symmetry of the 
 light-cone  superstring spectrum   
as its manifest  symmetry 
 we conjecture 
that it should also have a hidden fermionic  symmetry that
effectively interchanges  bosons  and fermions
and which  should  guide us towards understanding   the  relation 
between the two S-matrices. 
\end{footnotesize}
\end{abstract}
\newpage

\begin{small}
\vspace{2cm}
\tableofcontents
\end{small}
\setcounter{footnote}{0}
\renewcommand{\theequation}{1.\arabic{equation}}
\setcounter{equation}{0}
\newpage
\section{Introduction}

In this paper we continue the investigation of 
a particular 2-d massive integrable  theory which is 
the Pohlmeyer reduction of the \adss superstring theory  \cite{gt1,ms,gt2,rtfin,hit}.
The motivation  is to use the 2-d Lorentz-invariant 
reduced theory as a starting point for a first-principles solution of the 
 \adss superstring.

The original Pohlmeyer reduction  \cite{pol}  related the classical equations of motion of the $S^2$
sigma model  to the sine-Gordon equation. 
 In the  string-theory analog  of this reduction  \cite{tse,mi1}  one  considers the
    string  on $\mbb{R}_t \x S^2$ in the conformal gauge,  
   fixes  the residual conformal diffeomorphisms
   by the condition $t=\mu\tau$ and solves the Virasoro constraints in terms of
    one remaining degree of freedom,
    which is then interpreted as the sine-Gordon field.
     It is possible to extend this procedure to larger symmetric spaces  such as  $S^n$ and $AdS_n$
      and then further to the full superstring theory on $AdS_5 \x S^5$
      (see \ci{sssg,gt1,mira} for details  and references). 
 
 This reduction  
  associates  to  a classical string theory on a  coset  space $F/G$
  a 
classically equivalent  ``reduced'' theory. In general, this reduced theory can be described as 
a massive deformation  of a  gauged  $G/H$   WZW model (with an integrable potential 
proportional to $\mu^2$), which is 
related to a non-abelian Toda-type   generalization of  the sine-Gordon model.
The   number of independent fields 
 in the reduced  theory is the same 
as  the number of physical degrees of freedom of the 
 original string theory (with the conformal symmetry  completely fixed and $\mu$ playing 
the role of a fiducial conformal scale, i.e. an analog of $p^+$ in a light-cone gauge).
In addition, one   has the advantage of  manifest   2-d  Lorentz symmetry in the reduced theory, 
which is not usually present in a light-cone gauge fixed string theory  in   curved target space. 

This  is achieved at the expense of a non-trivial transformation,  essentially from coset 
coordinates to coset currents,  that  relates the fields of the original  superstring theory to the fields of the 
reduced theory. A precise formulation of the relation  between the two theories
(beyond  their classical equivalence including the correspondence  between  their integrable structures)  is therefore a   
non-trivial open question. 
 Since the   Pohlmeyer reduction utilizes the  classical conformal invariance, 
 it has a chance to continue to apply at 
the quantum level only if the conformal-gauge string sigma model one starts  with  is UV finite.
 This  is indeed the  case  for the  $AdS_5\x S^5$  
  superstring sigma model \ci{mt} (see \cite{dgt,rtt} and references therein). 
For consistency, the  corresponding  reduced theory  \ci{gt1,ms}
should   also be  UV finite, i.e. should have only a single 
 ``built-in'' scale $\mu$. 
 The UV finiteness  of the reduced theory  was    shown to be   the case at least   to the 
 two-loop order  and is expected to be true   to all  loop orders \ci{rtfin}. 
 Furthermore, it  was conjectured in $\cite{hit}$   that 
 the quantum partition functions of the string theory and the reduced  theory
  should  be  equal and 
 evidence  for that was provided  at the leading one-loop level.

\

To gain  better  understanding of the reduced theory  with the  eventual goal of 
 finding   its exact  solution 
  here we will  study the tree-level  S-matrix  for elementary excitations 
  of   this  theory   expanded near the  trivial vacuum. The spectrum of elementary excitations 
  contains 8   bosonic   and 8 fermionic 2-d degrees of freedom  of  the same mass  $\mu$ \ci{gt1}.
   This is  the same   spectrum as  found in the  $S^5$ choice of the  light-cone gauge 
 in the \adss  superstring theory \ci{mett,bmn},  but in contrast to the
  light-cone gauge fixed superstring action \ci{call,af,review}
   the interaction terms in the reduced action  are  2-d Lorentz invariant. 
   
 As a result, the  tree-level two-particle S-matrix of the reduced theory  that we will 
explicitly  compute below  is simpler  than its  counterpart  found 
directly   from  the superstring action \ci{kmrz}: while 
the two S-matrices turn out to have the same index  structure,  the  former depends 
only  on the difference of the two rapidities while  the latter
depends on both rapidities. 

The light-cone gauge \adss superstring S-matrix  (corresponding to the spin-chain 
magnon  S-matrix  on the gauge theory side \ci{stau}) 
plays an  important role 
in the conjectured Bethe Ansatz solution  for the superstring energy  spectrum 
 based on its  integrability 
   and  implied  by the 
AdS/CFT correspondence,  see \ci{review} for a review and further references. 
Its structure  is essentially fixed (up to a phase) by the residual global 
 $PSU(2|2) \times PSU(2|2) $ symmetry of light-cone gauge Hamiltonian
  \cite{bsmat,afz,review}.\foot{The tree-level superstring 
 S-matrix   was  shown to agree \cite{kmrz,afz} with a 
suitable limit of the  full S-matrix.}
However,  lack of 2-d Lorentz symmetry 
 leads to  complicated structure of the  corresponding thermodynamic Bethe  Ansatz 
 for the full quantum superstring spectrum 
 (see, e.g.,  \ci{aft,kaz}  and references therein). 
 
 The solution of the reduced theory is   expected to have a simpler form. 
Making a natural assumption   that  the classical integrability of the reduced theory 
extends to the quantum level,  its solution should 
be  determined, as in other similar \ci{p,mh,hsg} 
2-d  integrable massive QFT  examples \ci{zz,smatrev,hanran,mirp,cast,fri,braz,coaphd,brz}, 
 by an exact Lorentz-invariant   S-matrix. 

\

Let  us briefly recall the  structure of the  reduced theory action for the 
\adss superstring (see \ci{gt1,rtfin} for details). 
The starting point  is  the  set of  first-order  equations of motion  of 
 the $AdS_5 \x S^5$ superstring  \cite{mt,bb,rs,bpr}
  in the conformal gauge  written in  terms of the currents for 
the supercoset
\be \la{coo}
\fr{\hat{F}}{G}=\fr{PSU(2,2|4)}{Sp(2,2)\x Sp(4)}\,.
\ee
One can solve the Virasoro conditions  by  choosing a 
 particular constant  matrix  $T$  
from   the bosonic part of the coset algebra 
 and introducing a new set of variables, i.e. fields of the reduced theory, which are   algebraically
 related to the supercoset currents. Gauge-fixing the $\kappa$-symmetry, one can derive 
 the remaining independent equations of motion from a local action -- the reduced theory action. 
 The latter  
  happens to be the action of a  gauged WZW model for 
\be\la{co}
\fr{G}{H}=\fr{Sp(2,2)}{SU(2)\x SU(2)}\x\fr{Sp(4)}{SU(2)\x SU(2)}\,,
\ee
supplemented with an integrable bosonic potential and 
 fermionic terms. 
 The action of the gauge group $H=[SU(2)]^4$ commutes with the matrix $T$ so that the 
 $T$-dependent potential is also   gauge invariant. 
Explicitly, the  reduced  theory action  is given  by\foot{As discussed 
in \ci{rtfin}, it is natural to  assume that 
the overall constant $k$ here is  proportional to
the superstring tension, $\fr{\sqrt{\lambda}}{2 \pi}$. The reason is  that 
 the bosonic and the fermionic potential terms in this action  are directly related to the 
bosonic and the fermionic current terms  in the superstring action.
 This identification would make 
sense at the quantum level provided that  $k$ is not actually quantized  in the present 
case.}
\bea
{S} 
  =&k&\  \trm{STr} \Big[ \int d^2x \; {\textstyle{\fr{1}{2}}}  \ \gi\dpl g\ \gi\dm g\  -\,\int d^3x \;
   {\textstyle{\fr{1}{6}}}\ \epsilon^{mnl}  \ \gi \del_m g\ \gi\del_n g \ \gi\del_l g  \no
    \\
    &&\;\;\;\;\;+\;\int d^2x \  \big( A_+\dm g\gi-A_-\gi\dpl g-\gi A_+g A_-  +  A_+A_- 
     +\mu^2 \ \gi T g T\big)
\no 
\\
&&\;\;\;\;\;+\;\int d^2x \ \big(\psl T D_+ \psl+\psr T D_-\psr + \mu\; g^{-1}\psl g\psr\big)\Big]\,.
\la{act}
\eea
The fields in \rf{act} may be  represented 
by    $8 \x 8$  supermatrices  in the fundamental representation of $PSU(2,2|4)$ 
(with  diagonal $4\x4$ blocks 
being  bosonic and the off-diagonal $4\x4$ blocks being  fermionic).
 $g$ takes values in  $G=Sp(2,2)\x Sp(4)$ and $A_\pm$ in the  algebra 
   $\mf{h}$ of $H=[SU(2)]^4$. The fermionic  fields $ \psl,\psr$  originate from 
   particular components of the fermionic 
$\mf{psu}(2,2|4)$ superstring currents. 
   

\

In this paper we  shall present  the   reduced-theory counterpart of the  computation
   of the 
leading term in the  light-cone superstring S-matrix done in \ci{kmrz}: 
we shall expand the action \rf{act} around the  trivial 
 vacuum $g=\id,\,A_\pm=0,$ $\psr=\psl=0$  and find the  tree-level 
 two-particle scattering amplitude for the corresponding 8+8 massive elementary excitations. 
 To obtain   the effective quartic  Lagrangian   that 
 determines  this  amplitude   we shall choose the ``light-cone''  gauge $A_+=0$  
 (which preserves the  Lorentz symmetry in 2-d), 
 set   $g= e^\eta $ and expand 
 in powers of $\eta$. Splitting $\eta= \M+\xi$,  where $\M$ is from the  coset part of 
 $\mf{sp}(2,2)\oplus\mf{sp}(4)$  and $\xi$ is from the algebra of $H$, 
  and  solving the constraint that 
 follows from integrating out $A_-$  we will end up with the  following equivalent  Lorentz-invariant 
 Lagrangian  for the remaining 8+8 physical  fields, $\M$ and  $\psr,\psl$\foot{Here 
 $T=\fr{ i}{ 2} {\rm diag} (1,1,-1,-1,1,1,-1,-1)$ is a constant matrix.}
\be\la{spl}
\begin{split}
{L}_4= k\;\;&\trm{STr}\Big(\hs{3pt}\frac{1}{2}\dpl \M\dm \M-\frac{\mu^2}{2}\M^2+\psl T\dpl\psl+\psr T\dm \psr+\mu\psl\psr
\\& \quad\hs{5pt}+\fr{1}{12}\com{\M}{\dpl \M}\com{\M}{\dm \M}+\frac{\mu^2}{24}\com{\M}{\com{\M}{T}}^2
\\& \quad\hs{5pt}-\fr{1}{4}\com{ \psl T}{\psl}\com{\M}{\dpl \M}-\fr{1}{4}\com{\psr}{ T \psr}\com{\M}{\dm \M}-\frac{\mu}{2}\com{\M}{\psr}\com{\M}{\psl}
\\& \quad\hs{5pt} +\fr{1}{2}\com{ \psl T}{\psl}\com{\psr}{ T \psr}\Big)\,.
\end{split}
\ee
The  tree-level two-particle S-matrix  then follows directly from  the quartic terms in \rf{spl}.
The classical integrability \ci{gt1} of the reduced theory \rf{act} implies  
that the full  tree-level  S-matrix for the elementary excitations is determined by  the  
two-particle S-matrix  using   factorization. 

We shall compare the reduced-theory S-matrix   following from \rf{spl} to the 
light-cone  superstring one found in \ci{kmrz}. The two S-matrices 
 represent  the scattering of equivalent sets
 of degrees of freedom  and 
turn out to have the same  index  structure 
but   different kinematic coefficients.
One  important  outstanding question is if they  are actually  related, e.g., by 
a momentum-dependent non-Lorentz   invariant  transformation.
 
\


To find the   exact   solution  of the reduced theory
one is to  prove its quantum integrability  and 
 determine the exact quantum mass spectrum  and  the
 exact non-perturbative S-matrix.   
While the   form of the dispersion relation in the  light-cone superstring 
S-matrix is dictated by the centrally extended 
 $\mf{psu}(2|2)\oplus\mf{psu}(2|2) $  global
symmetry algebra of the  light-cone Hamiltonian  \ci{bsmat,hm,afz,review}
it remains to be  understood  if 
the exact mass spectrum  of the reduced  theory is also  effectively controlled by symmetries, 
e.g., by  a hidden fermionic (super)symmetry,  
 or if one 
needs to resort to a study  of  the (semiclassical) solitonic
 spectrum as  in \ci{csd,mirpa,cam}.

Motivated by  analogy with  similar models in \ci{hsg,mirp,cam},
the   hope for solving a theory like \rf{act} is based on the  possibility of using methods of 
deformed coset CFT, i.e. on the expectation that implications of 
conformal symmetry should survive the $\mu$-deformation 
and allow one to find the exact S-matrix. 
 Indeed, the bosonic part of the reduced theory action \rf{act} is  a special case of a 
  class of massive integrable deformations of the gWZW model,  generalized symmetric space 
  sine-Gordon models or generalized non-abelian Toda-type  models \ci{naat},  
  considered in \ci{sssg,hsg} (see \ci{coaphd} for a review and references).
   Since the full  fermionic model \rf{act} is UV finite
   \ci{rtfin}, we may 
   expect some important simplifications compared to the purely bosonic cases.\foot{Some lessons
    may   be drawn from 
    analogy  with the  UV finite $(2,2)$ supersymmetric sine-Gordon model 
   \ci{susyn2}
which  happens to be equivalent to   the reduced theory  for the $AdS_2 \times S^2$  
 superstring \ci{gt1}. However, the  sine-Gordon model 
  has topological solitons  and 
 lacks   the important feature 
  of non-trivial sigma model metric  in the kinetic part of the  action that 
  is characteristic to  higher-dimensional models like the one associated to 
  $AdS_3 \times S^3$  or  $AdS_5 \times S^5$.} 
  It is likely to be possible  to prove quantum integrability  by identifying (as, e.g.,  in \ci{gal})
  a higher spin  conserved current
  and  using the bootstrap method to determine the exact S-matrix.

\

The structure of the rest of this paper is as follows. 
We shall start in section 2 with an  analysis  of 
the massive integrable deformation  of the $G/H$ gauged WZW   model   
represented  by the bosonic  part of \rf{act}.
We shall choose the  $A_+=0$  gauge,   derive the bosonic part of 
the quartic Lagrangian \rf{spl} and demonstrate  its classical integrability. 

In section 2.4 we  shall compute the corresponding two-particle S-matrix 
and then in section 2.5  consider two special cases $G/H\,=\,SO(N)/SO(N-1)$ and $G/H\,=\,SO(N-1,\,1)/SO(N-1)$
  representing  the reduced theories  for strings 
 on $\mbb{R}_t \x S^N$ and $AdS_N \x S^1$ respectively.
 
 In section 2.6  we shall discuss  a  generalization of the bosonic part of \rf{act} 
 to an asymmetrically gauged case \ci{bars,hsg,qphd}  as  the 
  reduction procedure does not, in general, select  a particular gauging of the WZW model
 \ci{gt1}. We  shall show  that  the corresponding   tree-level 
 S-matrix  for elementary excitations 
  does not, in fact,  depend 
 on a particular choice of the $\tau$-automorphism defining the asymmetric gauging. This  
 will be  illustrated  on the example of the complex sine-Gordon model
  and its T-dual. 

In section 3 we shall turn to the  complete  reduced theory \rf{act} for the \adss  superstring 
including the fermionic terms. Fixing the $A_+=0$ gauge  and integrating 
out $A_-$   we will get a non-local effective  Lagrangian  for the physical 
components $\M,\psl,\psr$  but we will show that  there is an equivalent local Lagrangian \rf{spl} 
that leads to the same S-matrix. In section 3.4 we shall  consider the special cases 
of the reduced Lagrangians   for the $AdS_2 \times S^2$ and $AdS_3 \times S^3$
superstring models and show that they agree  with the corresponding special cases of \rf{spl}.
 
In section 4 shall 
 we compute the tree-level two-particle S-matrix for the complete  reduced theory 
 following from the Lagrangian  \rf{spl}.  
  The   reduced theory S-matrix  turns out to be of the
 same group-factorizable form as in the superstring case 
 \cite{kmrz}, which suggests 
   that there may be a direct  relation between the two S-matrices.


In section 5 we  shall make some concluding remarks  and discuss open problems. 

Appendix A contains   some details of simplification of the  gauge-fixed actions.
In Appendix B  we list the generators of the  relevant parts of the $\psu$  algebra. 
In Appendix C we give  explicit form of the $\mbb{T}$-matrix  of the  reduced theory
found in section 4. 
In Appendix D we consider the  tree-level two-particle S-matrix  of a (non-integrable) 
 massive  deformation 
of the bosonic geometric $G/H$ coset model and compare it to  the S-matrix of the 
 deformation of the gauged WZW model discussed in section 2.


\renewcommand{\theequation}{2.\arabic{equation}}
\setcounter{equation}{0}

\section{Tree-level perturbative S-matrix  of  bosonic\\  gauged WZW model
 with an integrable potential\la{smb}}
 
 Before turning to the superstring case we 
  shall start by considering  a  bosonic $G/H$  gauged WZW model with an integrable potential
 that appears \ci{sssg,gt1,mira} in the  Pohlmeyer reduction of the 
 geometrical $F/G$ coset sigma model,\foot{The action of the latter is 
 $S= -\ha k \int d^2 x\ \tr [( f^{-1} \del_+ f + {\cal A}_+)( f^{-1} \del_- f + {\cal A}_-)]$,
 where $f \in F, \ {\cal A} \in \mf{g}={\rm alg} (G)$.}
 or, equivalently, is associated to a string in the conformal gauge 
  moving on $\mbb{R}_t \times F/G$ space. 
To carry out the reduction one writes the equations 
of motion for the $F/G$ model in the  first-order form, i.e. in  terms of  currents, 
 and then solves the Virasoro conditions by introducing a scale $\mu$,  via 
 $t=\mu \tau$, a new field  $g \in  G$ and 
  a constant matrix $T$ in the 
   coset part $\mf{p}= \mf{f} \ominus  \mf{g}$
    of the algebra  $\mf{f}$ of $F$.
$H$  is  then the   subgroup of $G$ whose algebra commutes with $T$
    and the resulting classically equivalent integrable  theory  
    is described by a $G/H$ gWZW model with a potential determined by $T$, i.e. by 
    the bosonic part of \rf{act}. The fields of the reduced theory are 
     related to the currents of the original $F/G$ coset model  \ci{gt1}. 
    Examples of such theories have been  discussed, e.g.,  in \cite{p,mh,hsg}.
 
  The excitations around the trivial  $g=\id$ 
  vacuum of this theory are massive (with mass $\mu$) and  
  below we shall compute the corresponding 
  tree-level two-particle  S-matrix. Since the theory is classically integrable, 
  this then determines  the full tree-level S-matrix for elementary excitations 
  via factorization.


\subsection{Setup and notation\label{setup}}

Since we shall view the   $G/H$  gWZW model as a  reduced theory 
for the $F/G$ coset model it is natural to think of $G$ as being embedded into a group  $F$.  
Thus we  shall 
consider the three groups $F \supset G \supset H$, where $F/G$ and  $G/H$ are both symmetric coset spaces. The coset part of the algebra of $F$  will be denoted as   
 $\mf{p}$, i.e. $\mf{f}=\mf{g}\oplus\mf{p}$. We shall denote
  the maximal abelian subalgebra of 
 $\mf{p}$ as $\mf{a}$ and define $\mf{n}$ as the orthogonal complement of $\mf{a}$ in $\mf{p}$, i.e.
 \be \la{aa}
 \mf{f}=\mf{g}\oplus\mf{p} \ , \ \ \ \ \ \ \ \mf{p}=\mf{a}\oplus\mf{n} \ . \ee
 We will assume that $\mf{a}$ is one-dimensional and denote its non-trivial 
 element as $T$. Then 
 $H$ is defined as the subgroup of $G$ whose algebra $\mf{h}$  is 
  the centralizer of $T$ in  $\mf{g}$, i.e. $[T, \epsilon]=0, \ \epsilon \in \mf{h}$. 
  Then 
  \be \la{aac}
 \mf{g}=\mf{m}\oplus\mf{h} \ , \ \ \ \ 
\com{\mf{a}}{\mf{a}}=0\,, \ \ \com{\mf{a}}{\mf{h}}=0\,, \ \ \  \com{\mf{m}}{\mf{m}}\subset\mf{h}\ , 
\ \  \ \com{\mf{m}}{\mf{h}}\subset\mf{m}\,, \ \ \com{\mf{h}}{\mf{h}}\subset\mf{h}\,. \ee
We shall also assume the following commutation relations 
 which are all consistent with  $F/G$ being  a symmetric  space,
\be\ba{ccccc}
\com{\mf{a}}{\mf{n}}\subset{\mf{m}}\,,&\com{\mf{a}}{\mf{m}}\subset\mf{n}\,,&\com{\mf{n}}{\mf{n}}
\subset\mf{h}\,,&\com{\mf{n}}{\mf{m}}\subset \mf{a}\,,&\com{\mf{n}}{\mf{h}}\subset\mf{n}\,. \la{bcc}
\ea\ee
We shall consider  all the fields  as matrices with indices 
in the fundamental representation of 
 $\mf{f}$ or  $F$. The corresponding 
 orthonormal basis of generators of  $\mf{f}$ is   ($|\mf{f}|\equiv \dim \mf{f}$, 
 \ \ $\eta_{IJ}={\rm diag} ( \pm 1, ... \pm 1)$)
\be \la{orthog}
\com{T_I}{T_J}=f_{IJ}^{\;\;\;\;K}T_K \ , \ \ \ \ \ \ \ 
\STr(T_I T_J)=\eta_{IJ}\,,  \ \ \ \ \ \ 
I,J =0,\ldots, |\mf{f}| - 1\,.
\ee
We will use $\eta_{IJ}$ to raise and lower indices. 
Then  $f_{IJK}$ is totally antisymmetric.
Explicitly, the basis can be labelled  as follows:
\begin{itemize}
\item$T_0= n_0 T$: \  the   generator of  $\mf{a}$  \  \ ($|\mf{a}|=1$)
\item$T_a$:   basis of  $\mf{m}$  \  \ ($a=1,\ldots,|\mf{m}|$)
\item$T_i$:     basis of  $\mf{h}$ \ \ ($i=| \mf{m}|+1,\ldots,|\mf{g}|$)
\item$T_{\hat{a}}$:  basis of  $\mf{n}$\ \ ($\hat{a}=|\mf{g}|+1,\ldots, | \mf{f}|-1$) 
\end{itemize}
In this section  for simplicity we will  work with the assumption that $F$ is compact 
(and thus so are $G$ and $H$), 
 but generalization to a non-compact case will be straightforward. 
Compactness  implies that $\eta_{IJ}=-\delta_{IJ}$. 
Then $\STr(T^2_0)= - 1$ and thus 
  $n_0$  is a normalization constant entering the expression for $   \STr(T^2)$ 
\be\la{gamz}
\STr(T^2)= -n_0^{-2}\ . 
\ee
Also, 
for the basis of the $G/H$  part of the  algebra 
$\mf{m}$ we have 
\be\label{cond}
\STr\left(T_a T_b\right) =\eta_{ab}=- \delta_{ab}  \ .  
\ee
 We shall   assume that for $m \in \mf{m}$ 
\be\label{prop1}
\STr\ (\com{m}{T}^2 )=\STr\left(m^2\right)\,, \ \ \ \ \ \ {\rm i.e.}  \ \ \ 
n_0^{-2} f_{a\,0}^{\;\;\;\hat{a}}f_{b\,0\hat{a}} =-\delta_{ab} \ . 
\ee
This condition 
implies that the spectrum  of elementary excitations  
in the action like \eqref{act}  will be massive with the 
parameter $\mu$ being  the mass scale. 
We shall use  also 
the following  
 identity (below $m=m^aT_a \in\mf{m}$)
\be\la{identq}\ba{c}
\STr(\com{m}{\com{m}{T}}^2)= -n_0^{-2} f_{a\,0}^{\;\;\;\hat{a}}f_{b\hat{a}}^{\;\;\;0}f_{c\,0}^{\;\;\;\hat{b}}f_{d\hat{b}}^{\;\;\;0}m^a m^b m^c m^d = -n_0^{2}\, m^a m_a m^b m_b \ . 
\ea\ee



For  a non-compact group $G$, with $H$  being its 
maximal compact subgroup,   the sign of $\delta_{ab}$ in \rf{cond}  
and the overall sign in the action \rf{bact} below
  should be reversed.
 This prescription is consistent with the use of the supertrace 
 STr in \rf{act} and ensures
  that the propagating degrees of freedom enter the action with physical signs.
  In the non-compact case we will still
   demand \rf{prop1}, but  if $\STr(T_0^2)$ will change sign one will need to reverse the sign in 
    \rf{identq}  as well.

\subsection{Expansion of action near $g=\id$ vacuum \la{asymp}}

Our starting point is thus 
 the (symmetrically) gauged $G/H$ WZW model  with an integrable potential
\bea
&&\!\!\!\!\!\!\!\!\!\!\!\!\!\!\!\!{S}_{B}=-\,k\ \STr \Big[\,\int \left.d^2x \; \textstyle{\fr{1}{2}}  
\left(\gi\dpl g\ \gi\dm g\right)\right.-\; \int  d^3 x \; \textstyle{\fr{1}{6}}
\epsilon^{mnl}\left(\gi \del_m g\gi\del_n g\gi\del_l g\right)\no\\
&&\;
\;\;\;\;+\,\int d^2x \; \left( A_+\dm g\gi-A_-\gi\dpl g-\gi A_+g A_-  +  A_+A_-
 + \mu^2 \;\ \gi T g T\right) \Big]\,, \la{bact}
\eea
where $g\in G$, $A_\pm \in \mf{h}$ and $T$ is a constant matrix such that $\com{T}{H}=0$. The action is invariant under the following gauge transformations
\be\la{gag}
g\ra h^{-1}g h\  ,  \ \ \ \ \ \ \   A_\pm \ra h^{-1}A_\pm h+h^{-1}\del_\pm h\ , 
\ \ \ \ \ \ \ \ \ h=h(x)  \in H  \ . 
\ee
The corresponding equations of motion  imply
\be\la{flat}
\dpl A_- - \dm A_+ +\com{A_+}{A_-}=0\,,
\ee
that is the connection $A$ is flat, thus it can be gauged away on-shell. 
The remaining  equations of motion  for $g$ are then those of a non-abelian 
Toda model \cite{naat,sssg}
\be\la{eoma}\ba{c}
\dm(\gi \dpl g)=\mu^2\com{T}{\gi T g}\,,
\\(\gi \dpl g)_{\mf{h}}=0 \ , \ \ \ \ \ \ \ \ ( \dm  g \gi)_{\mf{h}}=0\,, 
\ea\ee
with  $g=\id$ is a  trivial 
vacuum point. 

To study the scattering of elementary excitations around this vacuum 
we shall set 
\be\la{lf1} g=e^\h\,, \ \ \ \ \ \ \ \  \h\in \mf{g}\,, \ee 
where  $\h \ra h^{-1} \h h$ under the gauge transformations \rf{gag}. 

Expanding the action 
 \eqref{bact}  in powers of $\h$ we get\foot{The expansion of the WZ term in the action can be determined 
 from the condition  of gauge invariance.}
\bea\la{lagex}
&& S_B= k \int d^2 x  \  \mathcal{L} \ , \ \ \ \ \ \ \ 
\mathcal{L}=\sum_{n=1}^{\infty}\;\mathcal{L}^{(n)}\,,\\
&&\Lag =-\STr\bigg(   \com{D_+}{D_-}\h + \sum_{n=1}^\infty  \fr{1}{\left(n+1\right)!}\Big[D_+\h\,\deta{n-1}{D_-\h}-\mu^2 \deta{}{T} \,\deta{n-1}{\deta{}{T}}\Big]\bigg) \no\\
&&\;\;\;\,=-\STr\Big[\com{D_+}{D_-}\h + D_+\h \,\fr{e^{\detaop}-1-\detaop}{\detaop^2}\,\left({D_-\h}\right)-\mu^2 \deta{}{T} \,\fr{e^{\detaop}-1-\detaop}{\detaop^2}\,\left(\deta{}{T}\right)\Big],\no
\eea
or explicitly,  
\bea 
&&		\Lag^{(1)} =-\STr\left(\com{D_+}{D_-}\h\right)\,, \la{uno} 
\\  && \Lag^{(2)} =-\STr\Big[\fr{1}{2}D_+\h D_-\h-\fr{\mu^2}{2}\deta{}{T}\deta{}{T}\Big]\,, \la{due} 
\\  && \Lag^{(3)} =-\STr\Big[\fr{1}{6}D_+\h\, \deta{}{D_-\h}-\fr{\mu^2}{6}\deta{}{T}\deta{2}{T}\Big]\,,
\la{tres}
\\  &&
\Lag^{(4)} =-\STr\Big[\fr{1}{24}D_+\h\, \deta{2}{D_-\h}-\fr{\mu^2}{24}\deta{}{T}\deta{3}{T}\Big]\,,\;\ldots\,\la{quatro} 
\\
&& D_\pm...\equiv \del_\pm...+\com{A_\pm}{...}\,,\hs{18pt}\deta{}{...}\equiv 
\com{\h}{...}\,.\la{takk}
\eea
This expansion is manifestly  gauge-invariant, e.g.,  invariant under the infinitesimal gauge transformations
\be \la{gaq}
\de \h = [\eta,  \ep]\,,\hs{20pt}\de A_\pm = D_\pm\ep\,,\ \ \ \ \ 
\ \ \ep(x) \in \mf{h} \ . 
 \ee
To compute the two-particle tree-level S-matrix we will not need terms
more than  quartic in fields. 

Next, let us decompose the fluctuation field $\h$ 
into  a coset (``physical'')  and  a subgroup (``gauge'')
  parts  according to \rf{aac}, i.e. 
\be \label{decomp}\h=\M+\xi\,,\hs{5pt}\trm{  }\hs{5pt}\ \ 
 \M \in \mf{m}\,,\hs{15pt}\xi \in \mf{h}\,.\ee
Then $\deta{}{T}=\dM{}{T}\equiv [\M,T]$, etc. 
The quadratic part of the  Lagrangian \rf{due}, which should determine the asymptotic scattering states, then takes the form 
\be\la{quadlag}
\Lag^{(2)}=-\STr\Big(\fr{1}{2}\dpl \M \dm \M-\fr{\mu^2}{2}\M^2+\fr{1}{2}\dpl \xi \dm \xi+A_+ \dm \xi -A_- \dpl \xi\Big)\,.
\ee
The corresponding  equations of motion (cf. \rf{flat},\rf{eoma})
\be\la{eoml}\ba{c}
\dpl\dm \M +\mu^2 \M=0\,,
\\\dpl\dm \xi+\dm A_+-\dpl A_-=0\,, \ \ \ \ \ \ \ \ \ \dm \xi=\dpl \xi=0\,.
\ea\ee
imply that only $\M$ represents propagating degrees  of freedom.\foot{This 
agrees of course 
with the general counting of degrees of freedom.
 We  started off with a Lagrangian containing $\dim G+2\dim H$ fields $g, A_+,A_-$. The off-shell gauge freedom and the two on-shell 
  constraints  arising from varying the action with respect to $A_\pm$  remove $3 \times \dim H$ degrees of freedom. This then leaves us 
  with $\dim G - \dim H = \dim \mf{m}$ degrees of freedom.}
   The apparent  massless and ghost modes in the $\xi,\,A_+,\,A_-$ part of \rf{quadlag}   
  are not physical states of the theory.

\subsection{$A_+=0$ gauge and action for physical  degrees of freedom 
 \la{gf}}

To compute the S-matrix  for physical excitations (in the $G/H$ coset directions)  one needs to
fix  the $H$ gauge symmetry in \rf{lagex}.  The final result 
for the S-matrix should be of course gauge-independent. 
One  apparently obvious  choice is to fix a gauge on  $g$
 and then integrate out $A_\pm$ which enter the action only algebraically. 
 However, the resulting action for $g$  then happens to be singular when expanded 
 near $g=\id$ \ci{gt1}.\foot{Indeed,  trying to integrate out $A_\pm$ in \rf{lagex}
 would lead to inverse powers of $\eta$.
 The reason for this  is clear from  \eqref{bact}: if we set $g=\id$, the
  terms quadratic in $A_\pm$ vanish.}
  This singularity, however, is a  gauge artifact (it is absent, in particular, 
  at the level of equations of motion where one can gauge fix $A_\pm$ to zero). 
   
 To get a regular expansion near $g=\id$ on should choose a gauge on 
 $A_\pm$. A very natural choice is the  ``light-cone'' gauge,  $A_+=0$.
 In addition to being  ghost-free (the ghost determinant is field-independent), in  2 dimensions 
  it also  preserves the Lorentz invariance of  the gauge-fixed  action
  found by setting $A_+=0$ in  \rf{uno}-\rf{quatro}.
  The quadratic Lagrangian is given by \eqref{quadlag} with $A_+=0$.
  After using integration by parts and  the cyclicity of the trace 
   the cubic and the quartic terms  take the following gauge-fixed form 
\be\label{cub} \begin{split}
\Lag^{(3)}_{g.f.}&=-\STr\Big(\frac{1}{2}\xi\com{\dm \M}{\dpl \M}+\frac{1}{6}\dpl \xi\com{\xi}{\dm \xi}+\frac{1}{2}A_-\left(\com{\M}{\dpl \M}+\com{\xi}{\dpl \xi}\right)\Big)\,,
\end{split}\ee
\be\label{quart} \begin{split}
\Lag^{(4)}_{g.f.}&=-\STr\Big(-\fr{1}{24}\com{\M}{\dpl \M}\com{\M}{\dm \M}+\frac{\mu^2}{24}\com{\M}{\com{\M}{T}}^2 \Big)
\\   &\hs{50pt}+\mathcal{O}\left(\M^2 \xi^2\right)+\mathcal{O}\left(\xi^4\right)+\mathcal{O}\left(A_-\M^2 \xi\right)+\mathcal{O}\left(A_- \xi^3\right)\,.
\end{split}\ee
We will not need $\xi$-dependent terms in \rf{quart} 
to determine the tree-level two-particle S-matrix for the physical 
states $\M$.

To explicitly decouple the unphysical degrees of freedom we may  integrate out $A_-$ 
and then solve the resulting delta-function 
constraint  expressing $\xi$ in terms of $\M$.\foot{An alternative 
approach (similar to the one in the standard  covariant gauge  choice) 
would be 
to  replace $A_-$ by $\del_- b$  and to treat $b, \xi, \M$ 
as a new set of fields with only $\M$  representing  asymptotic states. 
The resulting propagator for $(b,\xi)$ will 
have a ghost direction but that should not affect the unitarity of the final S-matrix for $\M$ (cf.  Lorentz gauge choice).}  
Integrating over $A_-$ in the sum of \rf{quadlag},\rf{cub} gives 
$(g^{-1} \del_+ g)_{\mf{h}}=0$, i.e. 
\be\la{cons}
\dpl \xi - \fr{1}{2}\com{\M}{\dpl \M}- \fr{1}{2}\com{\xi}{\dpl \xi}+...=0\,,
\ee
where dots 
 stands for higher order terms. Solving  for $\xi$ we find 
\be\label{xi}
\xi=\frac{1}{2}\fr{1}{\dpl}\com{\M}{\dpl \M}+\mathcal{O}(\M^3)\,.
\ee
The  lowest order 
is quadratic in $\M$, which means that to find 
the  two-particle S-matrix for $\M$ 
we do not need to consider any higher order terms in
 \eqref{xi}.\foot{This is also the reason why we do not need to consider the terms 
 neglected in \eqref{quart} and  can ignore the term of the form $\mathcal{O}\left(\xi^3\right)$ in \eqref{cub}.}

The resulting effective   Lagrangian for the physical degrees of freedom 
 $\M$ takes the form 
 $\Lag_{\M} = \Lag_{4\M} +\mathcal{O}(\M^5)$ 
 where 
\be\begin{split}\la{l1}
\Lag_{4\M}=&-\STr\left(\fr{1}{2}\dpl \M \dm \M-\fr{\mu^2}{2}\M^2+\fr{1}{8}\com{\M}{\dpl \M} \frac{\dm}{\dpl}\left(\com{\M}{\dpl \M}\right)\right.
\\&\;\;\;\;\;\;\;+\frac{1}{4}\com{\dm \M}{\dpl \M}\fr{1}{\dpl}\com{\M}{\dpl \M}
\\&\left.\;\;\;\;\;\;\;-\fr{1}{24}\com{\M}{\dpl \M}\com{\M}{\dm \M}+\frac{\mu^2}{24}\com{\M}{\com{\M}{T}}^2\right) 
\,.\end{split}
\ee
The first line in the above expression comes from 
\eqref{quadlag}, the middle line from \eqref{cub}, and the third line from \eqref{quart}. 
Using integration by parts (see Appendix A) 
eq.\eqref{l1} can be transformed into  the following local Lagrangian
\be\begin{split}\la{l2}
\Lag_{4\M}=&-\STr\Big(\fr{1}{2}\dpl \M \dm \M-\fr{\mu^2}{2}\M^2+\fr{1}{12}\com{\M}{\dpl \M}\com{\M}{\dm \M}+\frac{\mu^2}{24}\com{\M}{\com{\M}{T}}^2\Big) \,.
\end{split}
\ee
In general,  higher-order $\M^5$, etc.   terms in the Lagrangian \rf{l1}
may contain non-local factors $\fr{1}{ \del_+}$   (familiar from 
light-cone gauge fixed gauge theory in 4-d)\foot{As in 4-d gauge  
theory  such factors  should not cause problems with unitarity:
the original theory  we started with is  unitary 
for an appropriate choice of $G$ and $H$.}
but we expect\foot{An  indication that a  local action for the coset degrees of freedom should exist is that in the case 
where the $H$  gauge symmetry is fixed by a gauge on $g$ 
and $A_\pm$ are integrated out one gets a local 
(but degenerate  when expanded  near $g=\id$) action.
}  that,  
as happened  at quartic order, 
 they should effectively disappear (possibly after a field redefinition) 
and the end result should be a local 
Lagrangian for $\M$ only.

Let us now make    two remarks.
First, let  us comment on the  residual global symmetry of the resulting 
Lagrangian  for $\M$. As follows from \rf{gaq}  and \rf{aac}, 
splitting $\h= \M + \xi$  leads to the gauge transformations 
$\de \M = [\M,  \ep], \ \ \de \xi = [\xi,  \ep]$.
Since the 
residual  gauge transformations that preserve the  
condition $A_+=0$ are parametrized   by $\ep=\ep(x^-) \in \mf{h}$
it then follows that the effective  Lagrangian 
for $\M$ obtained by integrating out $A_-$ and $\xi$ should 
be invariant under $\de \M = [\M,  \ep]$, i.e. 
should  have at least  global $H$ invariance 
\be \la{he}
\M \to  h^{-1} \M h  \ , \ \ \ \ \ \ \ \ \ \   h=\const  \in H    \  . \ee
The corresponding S-matrix should then also 
have  global $H$  symmetry.

The second comment is about   integrability of the action  for $\M$.
The theory \rf{bact} we started with is integrable, i.e. 
its equations of motion follow from a flatness  condition
of the Lax connection (here $z$ is a spectral parameter) 
\be\la{lax}
\omega = d x^+\left(g^{-1}\dpl g+g^{-1}A_+g+z\mu  T\right)+dx^-\left(A_-+z^{-1}\mu g^{-1}Tg\right)\,.
\ee
The theory for the physical excitations $\M$  
obtained by eliminating the gauge degrees of freedom should then  
 also to be 
 integrable. Indeed, one can check that the equations
  of motion for $\M$ following from \eqref{l2}\foot{Here we have solved perturbatively for 
 $\del_+ \del_- \M$ and used the Jacobi identity
$\com{\com{\M}{T}}{\com{\M}{\com{\M}{T}}}=-\com{\com{T}{\com{\M}{\com{\M}{T}}}}{\M}-\com{\com{\com{\M}{\com{\M}{T}}}{\M}}{T}$ where 
 the first term on the right hand side vanishes. From
 \rf{aac},\rf{bcc} it follows  that $\com{\M}{\com{\M}{T}}\in \mf{a}$ 
 and thus the commutator of two such matrices should vanish 
 as   $\mf{a}$ is abelian.}
 \be\begin{split}\la{eomqm2}
\dpl\dm \M=-\mu^2 \M&+\fr{1}{6}\com{\dpl \M}{\com{\M}{\dm \M}}+\fr{1}{6}\com{\dm \M}{\com{\M}{\dpl \M}}\\&\hs{50pt}+\fr{\mu^2}{6}\com{T}{\com{\M}{\com{\M}{\com{\M}{T}}}}+\mathcal{O}(\M^4)\,,
\end{split}\ee
 can be obtained from the flatness condition of the following Lax connection
\bea
\omega_\M &=& d x^+\Big(\dpl \M-\fr{1}{4}\com{\M}{\dpl \M}-\fr{1}{12}\com{\M}{\com{\M}{\dpl \M}}+z\mu T\Big) 
 +dx^-\Big(\fr{1}{4}\com{\M}{\dm \M} \no \\
 &&\;\;\;+\;z^{-1}\mu\Big[ T-\com{\M}{T}+\fr{1}{2}\com{\M}{\com{\M}{T}}-
\fr{1}{6}\com{\M}{\com{\M}{\com{\M}{T}}}\Big] \Big)  +\mathcal{O}(\M^4)\,.
\la{laax}
\eea
 The  all-order completion of the Lagrangian 
 \rf{l2} should again admit a Lax connection.
 The integrability should then imply factorization of the tree-level 
 S-matrix for $\M$, i.e.  this S-matrix 
  should be essentially  determined by 
 the two-particle  scattering amplitude  following from 
 the simple quartic Lagrangian \rf{l2}.

 
\subsection{Tree-level two-particle S-matrix\la{tl1}}

Expanding the matrix field $X$ in the basis  of generators $\{T_a\}$ 
of the coset part of the algebra of $G$  and rescaling it 
by the overall coefficient $k$ (the expansion parameter) in the action \rf{bact}  
\be\la{fiedef}
\M= \fr{1}{\sqrt{k}} \M^aT_a\,,
\ee
we can write the quartic Lagrangian corresponding to \eqref{l2}, $\Lag'_{4\M}=k\,\Lag_{4\M}$, in component form
\be\la{lmgwzw}
\Lag'_{4\M}=
-\fr{1}{2}\dpl \M_a \dm \M^a+\fr{\mu^2}{2}\M_a \M^a+\fr{1}{12k}\gamma_{abcd}\M^a\M^c \dpl \M^b \dm \M^d+
\fr{1}{24k} \gamma_0 \mu^2 \M^a \M_a \M^b \M_b\,.
\ee
Here we used the relations \rf{orthog}--\rf{identq}  from section
 \ref{setup}\foot{Note that according to the definitions 
  in section \ref{setup} the indicies $a,\,b,\,c,\,d$ 
 should be raised and lowered with $\eta_{ab}=-\delta_{ab}$.  Thus the kinetic term in this Lagrangian has 
 the correct sign.}
 and defined  the coupling constants 
\be \la{cou}
\gamma_{abcd}\equiv  -f_{ab}^{\;\;\;i}f_{cdi} \ , \ \ \ \ \ \ \ \ 
\g_0 \equiv    n_0^2   \  . 
\ee
 Since there are no cubic terms in the Lagrangian there is only a single relevant tree-level
  Feynman diagram (Fig.1). 
\begin{figure}
\begin{center}
\epsfig{file=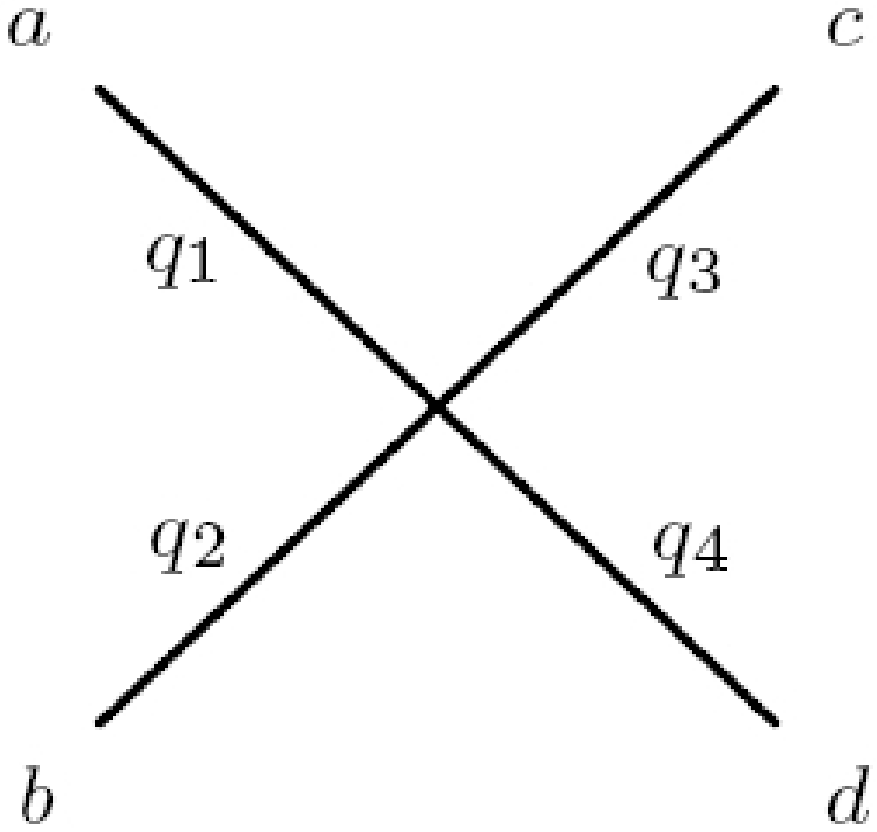, width=8cm}
\caption{Scattering diagram}
\label{ve}
\end{center}
\end{figure}
Using the operator approach, the usual expansion of the S-matrix
may be written as
\be\la{expansion}
\mbb{S}=1+\fr{i}{k} \mbb{T}\,,\ \  \ \ \ \ 
\mbb{T}=-k\int_{-\infty}^{\infty}d \tau\;\mbb{V}\left(\tau\right)+\ldots\,,  
 \ \ \ 
 \mbb{V}\left(\tau\right)=-\int_{-\infty}^{\infty}d\sigma \;\mathcal{L}_{int}\,,
\ee
which gives the following expression for the leading term in the $\mbb{T}$-matrix
\be\la{tb}
\mbb{T}= \fr{1}{12} \int d^2x \Big(\gamma_{abcd}\,:\M^a \M^c \dpl \M^b \dm \M^d:\,+ \fr{1}{2}\g_0 \mu^2 \delta_{ab}  \delta_{cd}\,:\M^a \M^b \M^c \M^d:\, \Big)\,,
\ee
 where $\M^a$ are now free-theory operators 
\bea
&&\M^a(x)=\int\fr{dp}{2\pi}\;\fr{1}{\sqrt{2 \epsilon}}\left(a_{p}^a e^{-i \Vec{p}\cdot \Vec{x}}+a_{p}^{a\dagger} e^{i \Vec{p}\cdot \Vec{x}}\right)\Big|_{\epsilon\,=\,\epsilon_p}\,,\la{md} \\
&&\com{\M^a(\sigma)}{\dot{\M}^b(\sigma')}=i\delta^{ab}\delta(\sigma-\sigma')\,,\hs{30pt}\com{a_{p}^a}{a_{p'}^{b\dagger}}=(2\pi)\delta^{ab}\delta(p-p')\,.
\la{comi}
\eea
We use the following notation
\be
\la{nott}
\Vec{x}=(\tau,\,\sigma)\ , \ \ \ \ \ \   \Vec{p}=(\epsilon,\,p) \ , \ \ \ \ \ 
x^\pm=\frac{1}{2} (\tau\pm\sigma)  \,,\hs{30pt} p_\pm=\epsilon\pm p\,, 
\ee
so  that $\Vec{p}\,\cdot \,\Vec{x} = p_+x^++p_-x^-$.
 We   denote the on-shell energy by $\epsilon_p$  and the rapidity by $\theta$:
\be\la{rap}
\epsilon_p =\sqrt{p^2+\mu^2} \ , \ \ \ \ \ \ 
\epsilon_p=\mu\cosh\theta \ , \ \ \ \  \ \ \   p=\mu\sinh \theta\,, \ \ \ \ \ \ 
p_\pm =\mu e^{\pm \theta}  \,.
\ee
We proceed by substituting $\eqref{md}$ into $\eqref{tb}$. 
As we are interested only in the action of $\mbb{T}$ on two-particle states we just consider the    
six terms that have two creation and two annihilation operators.  Integrating 
  over $\Vec{x}$ we end up with a  factor 
   $\delta^{(2)}(\Vec{q}_1+\Vec{q}_2-\Vec{q}_3-\Vec{q}_4)=\delta(\epsilon_{q_1}+\epsilon_{q_2}-\epsilon_{q_3}-\epsilon_{q_4})\delta(q_1+q_2-q_3-q_4)$. 
   These two $\delta$-functions can then be used to do the integral over the momenta 
   $q_3$ and $q_4$, as the arguments of the $\delta$-functions only vanish when either $
\Vec{q}_1=\Vec{q}_3 \hs{5pt} \trm{and}\hs{5pt}\Vec{q}_2=\Vec{q}_4\,,$
or $\Vec{q}_1=\Vec{q}_2 \hs{5pt} \trm{and}\hs{5pt}\Vec{q}_2=\Vec{q}_3.$
We also pick up a  Jacobian factor 
$
\frac{\epsilon_{q_1}\epsilon_{q_2}}{|\epsilon_{q_1}q_2-q_1\epsilon_{q_2}|}\,.
$
This leads to  
\be\begin{split}
\mbb{T}&=\int \fr{dq_1 dq_2}{(2\pi)^2}\fr{1}{4|\epsilon_{q_1}q_2-q_1\epsilon_{q_2}|}\Big[\Big(\fr{1}{12}\gamma_{abef}(2\mu^2-q_{1+}q_{2-}-q_{2+}q_{1-})
\\&\hs{70pt} +\fr{1}{12}\gamma_{aebf}(-4q_{1+}q_{2-})+\fr{1}{12}\gamma_{afbe}(-2\mu^2-q_{1+}q_{2-}-q_{2+}q_{1-})\Big)
\\&\hs{70pt}+\Big(\gamma_0\fr{\mu^2}{6}\delta_{ab}\delta_{ef}+\gamma_0\fr{\mu^2}
{6}\delta_{ae}\delta_{bf}+\gamma_0\fr{\mu^2}{6}\delta_{af}\delta_{be}\Big)\Big]a^{a\dagger}_{q_1}
a^{b\dagger}_{q_2}a^{e}_{q_1}a^{f}_{q_2}\, .
\end{split}
\ee
The action of $\tm$ on  two-particle states 
\be
\ket{\M^{c}(p_1)\M^{d}(p_2)}=
2\sqrt{\epsilon_{p_1}\epsilon_{p_2}}a_{p_1}^{c\dagger}a_{p_2}^{d\dagger}\ket{0}\,
\ee
gives  coefficient functions   $T_{ab}^{cd}(\vartheta)$
\be\la{vetd}
\mbb{T}\ket{\M^{c}(p_1)\M^{d}(p_2)}=T_{ab}^{cd}(\vt)\ket{\M^{a}(p_1)\M^{b}(p_2)}\,. 
\ee 
As usual in a 2-d Lorentz invariant theory, 
 $T_{ab}^{cd}$ should  depend on the  difference of the two 
 rapidities 
\be\la{vtd}
\vt \equiv \left|\theta_1-\theta_2\right|\,, \ \ \ \ \ \ \ \ \ \ 
p_{1\pm }=\mu e^{\pm \theta_1} \ , \ \ \ \ p_{2\pm }=\mu e^{\pm \theta_2} \ . 
\ee
Using  that
\bea
&&a^{a\dagger}_{q_1}a^{b\dagger}_{q_2}a^{e}_{q_1}a^{f}_{q_2}\ket{\M^{c}(p_1)\M^{d}(p_2)}
\no 
\\
&&=\left(2\pi\right)^2\sqrt{\epsilon_{p_1}\epsilon_{p_2}}
\Big(\delta^{ce}\delta^{df}\delta(q_1-p_1)\delta(q_2-p_2)+\delta^{cf}
\delta^{de}\delta(q_1-p_2)\delta(q_2-p_1)\Big)a^{a\dagger}_{q_1}a^{b\dagger}_{q_2}\ket{0}\,,
\no \eea
we end up with 
\be\la{t}\begin{split}
T_{ab}^{cd}(\vt)=&\fr{1}{12\sinh \vt}\Big[\gamma_0\left(\delta_{ab}\delta^{cd}+\delta_{a}^{c}\delta_{b}^{d}+\delta_{a}^{d}\delta_{b}^{c}\right) +\gamma_{ab}^{\;\;\;cd}-\gamma_{a\;\,b}^{\,\;d\;\,c}
\\&.\hs{50pt}-\left(\gamma_{ab}^{\;\;\;cd}+2\,\gamma_{a\;\,b}^{\,\;c\;\,d}+\gamma_{a\;\,b}^{\,\;d\;\,c}\right)\cosh \vt\Big]\,.
\end{split}
\ee

The same result is found by using the Feynman rules, i.e. 
the  LSZ reduction of the four-point vertex function in \rf{l2}. The 
 vertex function  corresponding to the  diagram in Figure 1  with 
 all the momenta $q_i$  flowing in is
\begin{equation}\la{fd}
\begin{split}
V_{abcd}=\fr{i}{12k} \Big(\gamma_{abcd}\ q_{2+}q_{4-}+\ha \gamma_0{\mu^2 } \delta_{ab}\delta_{cd} \Big)+ \trm{permutations}\,, \end{split}\end{equation}
were  there are  23 permutations of $\{(a,\,q_1),\, (b,\,q_2),\,(c,\,q_3),\,(d,\,q_4)\}$ other than the identity.
Using that all of the legs are on-shell, i.e.  
$
q_{1\pm }=-q_{3\pm }=\mu e^{\pm \theta_1}\,, 
\ \ 
q_{2\pm}=-q_{4\pm}=\mu e^{\pm \theta_2}\,
 $
we again  find \rf{t}, with
\be\la{rel}
T_{ab}^{cd}(\vt)=\fr{k}{4i\mu^2\sinh \vt}\;V_{ab}^{cd}(\vt)\ . \ee

\subsection{Examples: $SO(N)/SO(N-1)$ and $SO(N-1,1)/SO(N-1)$\la{sons}}

Let us now  consider two specific examples, where $G/H\,=\,SO(N)/SO(N-1)$ and $G/H\,=\,SO(N-1,\,1)/SO(N-1)$. The corresponding actions 
\rf{bact}   describe Pohlmeyer reduced theories  corresponding to 
 string theory on $\mbb{R}_t \x S^N$ and $AdS_N \x S^1$ respectively; 
 the $N=5$ case is thus  relevant for the $AdS_5 \x S^5$  superstring
 \ci{gt1}.
 Below we shall find  the explicit expressions 
 for  the coefficients  $T_{ab}^{cd}(\vt)$ in \eqref{rel},\rf{t}.
 
\

\noindent{
\, \;$\mbf{G/H\,=\,SO(N)/SO(N-1)}$}

\

\noindent
Here $F=SO(N+1), \ G= SO(N), \  H=  SO(N-1)$ (cf. section 2.1).
 The standard normalised basis for $\mf{f}=\so(N+1)$ in terms of 
  $(N+1) \x (N+1)$ matrices  is 
\be\la{qqq}
 T_I=\{\tilde{T}_{\al\bet}\ ,\  \al<\bet\}\,,\hs{20pt}\trm{ }\hs{10pt}(\tilde{T}_{\alpha\beta})_{uv}=\frac{1}{\sqrt{2}}\left(\delta_{\alpha u}\delta_{\beta v}-\delta_{\alpha v}\delta_{\beta u}\right)\,,\ee
where $\alpha,\,\beta,\,\ldots,\,u,\,v,\ldots=1,\,\ldots,\,N+1$.
As in  \cite{sssg,gt1} we may choose $\mf{g}$ to be  formed by $(N+1) \x (N+1)$ matrices 
with non-trivial lower right  $N \x N$ corner  and $\mf{h}$ 
formed by matrices with non-trivial lower right  $(N-1) \x (N-1)$ corner. This corresponds to choosing the basic matrix $T$ as 
\be\la{tt}
T = \sqrt{2}\,\tilde{T}_{01}=\left(\ba{ccccc}
0&1&0&\ldots&0
\\-1&0&0&\ldots&0
\\0&0&0&\ldots&0
\\\vdots&\vdots&\vdots&\ddots&\vdots
\\0&0&0&\ldots&0
\ea\right)\,.
\ee
Then  $n_0 = \fr{1}{\sqrt 2} $  in \rf{gamz} 
and  the condition \eqref{prop1} is satisfied. 
An orthonormal basis of $\mf{m}$ is  given by $T_a=\tilde{T}_{2,a+2}$, where $a=1,\,\ldots,\,N-1$.
The coupling constants defined in \rf{cou}  are found to be 
\be \gamma_{abcd}=\frac{1}{2}\left(\delta_{ac}\delta_{bd}-\delta_{ad}\delta_{bc}\right)\,, \ \ \ \ \ \ \ \ \    \g_0 =   \frac{1}{2}  \ . \la{cpp}     \ee
Using \eqref{t}  we see that $T_{ab}^{cd}$ in \rf{t} here 
simplifies to
\be
\la{son}
\begin{split}
T_{ab}^{cd}(\vt)=\fr{1}{8\sinh \vt}\Big[\delta_{a}^{c}\delta_{b}^{d}-\left(\delta_{ab}\delta^{cd}-\delta_{a}^{d}\delta_{b}^{c}\right)\cosh \vt\Big]\,.
\end{split}
\ee
When $N=2$ the subgroup $H$ is trivial and in this case 
the action \rf{bact} reduces to that of the  sine-Gordon model
\ci{sssg}. 
In this  case the $G/H$  coset is one dimensional  and thus \rf{son} becomes 
\be
\la{sg}
\begin{split}
T(\vt)=\fr{1}{8\sinh \vt}\,,
\end{split}
\ee
which agrees with the two-particle tree-level amplitude 
 for scattering of elementary  excitations around 
 trivial vacuum of  the sine-Gordon  model. 

For  $N=3$ the symmetrically gauged model \rf{bact} is related to the  complex sine-Gordon 
(CsG) model:
fixing the  $H=SO(2)$ gauge on $g$ and integrating out $A_\pm$ we end up with 
a model that is equivalent  to the CsG model by 2-d duality. 
The above procedure  based on the $A_+=0$ gauge  leads to the  S-matrix 
\eqref{son}. The latter is also  the same 
 as  the two-particle tree-level scattering amplitude for elementary excitations in
 CsG model. The reason for this relation has to do with the existence of several 
  possible $H$ gaugings of the WZW model and will be  discussed further in section \ref{tauaut}.

\

\noindent{
\,\;$\mbf{G/H\,=\,SO(N-1,1)/SO(N-1)}$}

\

\noindent
Here  $F=SO(N-1,\,2), \ G=SO(N-1,\,1), \ H=SO(N-1)$. As here   
 $G$ is non-compact and $H$ is its maximal compact subgroup, 
as discussed in section \ref{setup} the sign  in the relations in 
\rf{gamz},\rf{cond}, etc. (i.e. the sign of the trace) 
 should then be reversed. 
 The basis  for $\mf{f}=\so(N-1,2)$ can be chosen as
\be T_I=\{\tilde{T}_{\al\bet}\ ,\  \al<\bet\}\,,\hs{20pt}\trm{ }\hs{10pt}(\tilde{T}_{\alpha\beta})^{u}_{\;v}=\frac{1}{\sqrt{2}}\left(\delta_{\alpha}^{ u}\tilde{\eta}_{\beta v}-\tilde{\eta}_{\alpha v}\delta_{\beta}^{ u}\right)\,,\ee
where  
 $\tilde {\eta}=\trm{diag}\left(-1,-1,1,1,\ldots,1\right)$.
 Using the same embedding of $\mf{g}$ and $\mf{h}$ into $\mf{f}$ as in 
 the previous example, the choice of $T$ is then the  same as in \eqref{tt}. 

An orthonormal basis of the coset part of the algebra, 
$\mf{m}$, is  given by $T_a=\tilde{T}_{2,a+2}$, where $a=1,\,\ldots,\,N-1$. As 
the basis for $\mf{h}$ is the same as in the previous example it is clear that 
changing the sign of the trace  gives
\be \gamma_{abcd}=-\frac{1}{2}\left(\delta_{ac}\delta_{bd}-\delta_{ad}\delta_{bc}\right)\,,
 \ \ \ \ \ \ \ \ \    \gamma_0=-\frac{1}{2}  \ ,  \ee
instead of \rf{cpp}. 
   Using \eqref{t}  we see that here $T_{ab}^{cd}$ simplifies to
\be
\la{sonc}
\begin{split}
T_{ab}^{cd}(\vt)=-\fr{1}{8\sinh \vt}\Big[ \delta_{a}^{c}\delta_{b}^{d}-
 (\delta_{ab}\delta^{cd}-\delta_{a}^{d}\delta_{b}^{c})\cosh \vt\Big]\,.
\end{split}
\ee
This  differs from \eqref{son} only  by the overall  sign.


\subsection{Asymmetrically gauged WZW model with an integrable potential\la{tauaut}}

The action \rf{bact} we discussed above corresponds to the symmetrically gauged WZW 
model, but  there is a more general class of models 
with an asymmetrical gauging of $H$.
Asymmetrical gauging \cite{bars,hsg,qphd} uses 
   an anomaly-free automorphism $\tau$ of the algebra $\mathfrak{h}$
   (in the symmetrically gauged  case $\tau= {\id}$).\foot{Anomaly freedom is equivalent
    to   $\STr(\tau(a)\tau(b))=\STr(ab)$, for all $a,\,b\,\in\mf{h}$.} 
As discussed in  \cite{gt1,gt2,mira,hit}, 
 an asymmetrically  gauged analog of the $G/H$ 
   gWZW action in \eqref{bact}  may also be considered 
 as a Pohlmeyer reduction of the $F/G$ coset model. 
 

As we shall see below, the choice of the 
 $\tau$-automorphism does not actually affect the 
 perturbative S-matrix of the corresponding theory, i.e.  
  all the different anomaly-free  gaugings of the WZW model with a
   potential  have the same perturbative  S-matrix. 
 A way to see this is  by observing that all dependence on $\tau$ 
 in the action drops out in the $A_+=0$ gauge, so that 
  the spectrum of excitations 
 near the $g=\id$  vacuum  and the corresponding S-matrix 
 do not  depend on $\tau$.

The action of the asymmetrically gauged model is obtained   from  \eqref{bact}
by replacing the $A_+ A_-$ term by $\tau(A_+)A_-$, i.e. by replacing the $A_\pm$ dependent terms  in \eqref{bact}   by 
\be\la{asymact}
 S_{a.g.}= -k\;\int d^2x \; \STr\Big[A_+\dm g\gi-A_-\gi\dpl g-\gi A_+g A_-  +  \tau(A_+)A_-\Big]\,.
\ee
The full  action is then  invariant under the following gauge transformations
\be\la{asa}
g\ra h^{-1}g\hta(h)\,, \ \ \  A_+\ra h^{-1}A_+ h+h^{-1}\dpl h\,,\hs{10pt} A_- \ra \hta(h)^{-1}A_-\hta(h)+\hta(h)^{-1}\dm \hta(h)\,,
\ee
where $\hat{\tau}$ is the lift of the automorphism $\tau$ from the algebra $\mathfrak{h}$ 
to the group $H$.

The analog of  the quadratic Lagrangian before gauge-fixing \rf{quadlag}    is
\be\la{asw}
\Lag^{(2)}=-\STr\Big[\fr{1}{2}\dpl \M \dm \M-\fr{\mu^2}{2}\M^2+\fr{1}{2}\dpl \xi \dm \xi+A_+ \dm \xi -A_- \dpl \xi-A_+A_-+\tau(A_+)A_-\Big] \,,
\ee
leading to the following  linearised equations of motion
\bea  &&
\dpl\dm \M +\mu^2 \M=0\,, \ \ \ \ \ \ \ \ \dpl\dm \xi+\dm A_+-\dpl A_-=0\,,\no
\\
&&\dm \xi-A_-+\tau^{-1}(A_-)=0\,,\hs{20pt}\dpl \xi+A_+-\tau(A_+)=0\,.
\la{taueom} \eea
It is easy to see that it is possible again 
to set $A_\pm=0$ by combining 
 the equations of motion and the gauge freedom.\foot{For example, we may start 
  by fixing the  $A_+=0$ gauge.
  The residual gauge freedom (with parameter $\ep(x^-)$) 
  can be  fixed by demanding that $(A_--\dm\xi)(0,\, x^-)=0$. 
  From the second equation  we then have that $\dpl\left(\dm \xi- A_-\right)=0$, which  implies  $A_-=\dm\xi$ for all $x^+$.  The third equation   then implies that $\tau^{-1}(A_-)=0 \;\Ra\; A_-=0$.} 
Once  $A_\pm=0$,  it follows  from \rf{taueom} that  $\xi$ is non-propagating, 
i.e. the physical asymptotic states are again  those of the coset components $\M$.


Fixing the $A_+=0$ (or $A_-=0$) gauge in the asymmetrically gauged action 
containing \rf{asymact} we observe that  the dependence on the choice of 
$\tau$-automorphism disappears so we should  end up with the same (gauge-independent) 
S-matrix as in the symmetrically gauged case discussed above.

To give an explicit   example,    let us 
consider  the case with  $G/H=SO(3)/SO(2)$  embedded into  $F=SO(4)$
(cf.   section \ref{sons}). 
 Fixing the $H$-gauge  on  $g$  as  in \cite{gt1} we may  
 parametrize it in terms of two  coset  coordinates $\phi,u$ \foot{
 This is effectively a gauge choice on $\eta$ in \rf{decomp} 
 relating $\M$ and $\xi$.}
\be
g=e^{u T'_3}e^{2\vp T'_1}e^{ u T'_3}\,, \ \ \ \ \ \ \  \ \   T_I'= \sqrt 2  T_I \ , \la{geg}
\ee
where $T_1$ is one of two  coset generators and $T_3$ is a generator of $\mf{h}=\so(2)$
($T_I$ are defined as in \rf{qqq}). 
 Starting with the symmetrically gauged action \rf{bact}
 and integrating out $A_\pm$  we  get 
\be\la{csg2}
\tilde \Lag_{csg}= 4 \big( 
\dpl \vp \dm \vp + \cot^2  {\vp} \;\dpl  u \dm  u 
+ \fr{\mu^2}{2} \cos  2 {\vp} \big) 
\,.
\ee
The expansion of this Lagrangian near the trivial vacuum 
$g=\id$, i.e.  $\vp=u=0$  is obviously  singular.\foot{The singularity in the second term 
 is due to the degeneration of 
$- g^{-1} A_+ g A_-  + A_+ A_-$ term in \rf{bact} near $g=\id$.}
 For an abelian  $H$  the non-trivial choice of the  $\tau$-automorphism is $\tau(\mathfrak{h})=-\mathfrak{h},\  \hat \tau(h) = h^{-1}$ leading to the  
  ``axially'' gauged theory.  Starting from \rf{bact}
  with  $A_+ A_-$ replaced according to \rf{asymact} by $-A_+ A_-$, 
  choosing the condition  on $g$ 
  (now fixing the axial gauge symmetry $g \to h^{-1} g h^{-1} $)
  as 
  \be
g=e^{- u T'_3}e^{2\vp T'_1}e^{ u T'_3}\,, \la{geg2}
\ee
  and integrating out $A_\pm$ we then get, instead of  \rf{csg2}, 
    the complex sine-Gordon model 
\be\la{csg1}
\Lag_{csg}=
4\big( \dpl \vp \dm \vp + \tan^2 {\vp} \;\dpl  u \dm  u 
+ \fr{\mu^2}{ 2}  \cos 2  \vp \big) 
\,. 
\ee
This  Lagrangian  has a regular expansion near $\vp=u=0$, i.e. 
  in the asymmetrically gauged case the  choice of the gauge \rf{geg}
is non-singular for the purpose  of computing the S-matrix. 

One  can check directly that the   resulting  S-matrix for  \eqref{csg1}
is the   same as  found above 
in the $A_+=0$ gauge in the symmetrically gauged  case, 
in agreement with  the 
gauge-independence of the S-matrix and the above observation that  the 
S-matrix in the $A_+=0$ gauge should not depend on a choice of the  $\tau$-automorphism. 
Indeed, expanding \rf{csg1}  and 
introducing  the two ``cartesian'' coordinates $z_a$ 
with the standard normalization of the kinetic term 
via 
   $z_1+i z_2=  2 \sqrt 2 \vp\, e^{i u}$
we get  (cf. \rf{lmgwzw},\rf{cpp})\foot{
One may start of course directly from the 
 well-known alternative parametrization  of \eqref{csg1} $ 
\psi= 2\sqrt{2} \sin \, {\vp}  \,e^{iu}= z + O(z^3) $
where  $
\mathcal{L}_{csg}=\fr{1}{4}\fr{\partial_+\psi\partial_-\psi^*   +  \partial_-\psi\partial_+\psi^* }{1-\fr{1}{8}|\psi|^2}
-\fr{\mu^2}{2}|\psi|^2 \,.
$ 
The quartic expansion of this Lagrangian is related to \eqref{csge} by
 a local field redefinition that does not affect the S-matrix.}
\be\la{csge}
\Lag_{csg}=\fr{1}{2}\dpl z_a \dm z^a-\fr{\mu^2}{2}z_az^a +\fr{1}{24}(\delta_{ac}\delta_{bd}-\delta_{ad}\delta_{bc})z^a z^c \dpl z^b  \dm z^d+\fr{\mu^2}{48}z_a z^a z_b z^b+\mathcal{O}(z^5)\,.
\ee
This Lagrangian then leads to  the same tree-level two-particle S-matrix  \eqref{son} 
as the symmetrically gauged $SO(3)/SO(2)$ model.

Let us note that 
in  the case  when $H$ is abelian 
 the  sigma models obtained by integrating out $A_\pm$ 
 in the symmetric and asymmetric gauged actions 
 are   related  by the  T-duality (or the scalar-scalar 2-d duality). This is 
  evident, e.g.,  from comparing 
 \rf{csg2}   and   \rf{csg1}
 (see  also   \ci{mira1,gt1}).
 This duality, in general,  may  map fluctuations near 
 a trivial vacuum of one model into  fluctuations near a non-trivial 
 background  of its dual.

The perturbative  S-matrix  of the  complex sine-Gordon 
model was discussed in  \cite{csg}   and its exact  non-perturbative (solitonic) 
 generalisation in  \ci{csd}. 
In particular,  this  theory contains non-topological solitons 
whose small-charge limits are the elementary excitations near 
the trivial vacuum we discussed above.  Therefore,
 an appropriate  limit of the full  CsG 
S-matrix should agree with \eqref{son} for  $N=3$. 
Note that this case is   different from 
the  sine-Gordon model where 
the solitons are topological, i.e. 
 they interpolate between different vacua and thus  do not reduce 
to  elementary excitations in any limit.\foot{
While the vacuum structure of the two models  is formally 
the same, the kinetic term in CsG model prohibits solitons interpolating 
between two vacua.}

\renewcommand{\theequation}{3.\arabic{equation}}
\setcounter{equation}{0}

\section{Expansion  of  action  of 
 reduced form of \\  $AdS_5 \x S^5$ superstring model}


Let us now turn to  the fermionic extension \rf{act} 
of the gWZW model with integrable potential \rf{bact}
arising  from the Pohlmeyer reduction of the $AdS_5 \x S^5$ superstring sigma model
\ci{gt1,ms}. The fields in this  action  
are related to  the currents of the superstring  sigma model 
based on the supercoset $
\fr{PSU(2,2|4)}{Sp(2,2)\x Sp(4)}\,.
$
 We shall  use a particular matrix representation for these fields -- 
 the same as
  in \cite{gt1,rtfin,hit}. 
We shall start  with recalling the basic definitions and notation 
and then work out the  expansion of the action in components, 
generalizing \rf{lmgwzw} to include the  fermionic terms  in \rf{spl}.

\subsection{Setup   and   notation\la{setup2}}

The bosonic part of the supercoset $\fr{\hat{F}}{G}= \frac{PSU(2,2|4)}{Sp(2,2)\x Sp(4)}$ is $\fr{{F}}{G} = \frac{SU(2,2)\x SU(4)}{Sp(2,2)\x Sp(4)}$.
 The Pohlmeyer reduction of the bosonic part 
 produces  the direct sum of 
 the two   models of the type 
 discussed in section \ref{smb}: 
\be\la{gr}\ba{c}  F_{_{(1)}}=SU(2,2) \ , \ \ \ \ \ \ 
 \;\;\fr{G_{_{(1)}}}{ H_{_{(1)}}}=\frac{Sp(2,2)}{SU(2)\x SU(2)}\,, 
\\  F_{{(2)}}=SU(4) \ , \ \ \ \ \ \   \;\;\fr{G_{_{(2)}}}{ H_{_{(2)}}}=\frac{Sp(4)}{SU(2)\x SU(2)}\ .
\ea\ee
The fields of these  two models can be
described together as $4 \x 4$  blocks of  $8 \x 8$  
matrices  in  the fundamental  representation of  $\hat{F}=PSU(2,2|4)$ (see 
 \cite{gt1,rtfin,hit} for details). Schematically, $f \in PSU(2,2|4)$ is represented by 
\be\la{mats}
\left(\ba{cc}SU(2,2)& \ \trm{Grassmann}\ \\ \ \trm{Grassmann}\ &SU(4)\ea\right)\,. \ee 
The corresponding algebra $\hat{\mf{f}}=\psu$ admits a $\mathbb{Z}_4$ decomposition \cite{bb}
\be\la{fef}
\hat{\mf{f}}=\mf{f}_0\oplus\mf{f}_1\oplus\mf{f}_2\oplus\mf{f}_3,\,\hs{20pt}\com{\mf{f}_i}{\mf{f}_j}=\mf{f}_{i+j \trm{ mod } 4}\,,
\ee
where $\mf{f}_0$, $\mf{f}_2$ are bosonic and $\mf{f}_1$, $\mf{f}_3$ are fermionic. These
 subspaces relate to those of section \ref{smb} as follows:  $\mf{f}_0 =\mf{g}=\mf{sp}(2,2) 
 \oplus \mf{sp}(4)$
 and $\mf{f}_2=\mf{p}$ is the orthogonal complement of
  $\mf{g}$ in the bosonic  part  $\mf{f}=\su(2,2)\oplus \su(4)$     of the superalgebra, i.e.
 $\mf{p}$  is the  part of the algebra that corresponds to the coset space 
 $AdS_5 \times S^5$. 

The reduced  action \rf{act} is constructed using 
the supertrace
\be\la{stre}
 \trm{STr} (f)=\sum_{i=1}^4 f_{ii}-\sum_{i=5}^8 f_{ii}\,,
\ee
The matrix $T$ from the maximal abelian  subalgebra of $\mf{p}$ satisfying the
 required properties is chosen as 
in \ci{gt1}
\be\la{req}
T=\left(\ba{cc}T_{_{(1)}}&0\\0&T_{_{(2)}}\ea\right)\,,\ \ \ \ 
T_{{(1)}} = T_{{(2)}}=\fr{i}{2}\trm{diag}\;\left(1,\,1,\,-1,\,-1\right)\,,\hs{20pt} 
T^2=-\fr{1}{4} \mbf{I} \,.
\ee
With this choice $\STr ( T_{_{(1,2)}})^2 = -1$, i.e. $n_0=1$ in \rf{gamz}. 
As in section \ref{smb},
  the subgroup $H$ 
 of $G=Sp(2,2)\x Sp(4)$ has  the  algebra $\mf{h}$   defined to be the centralizer of $T$.
 Here  $H=SU(2)^4$, which is embedded into $8 \times 8$ 
 matrix  representation of $\hat{F}=PSU(2,2|4)$ as follows
\be\la{ssss}
\left(\ba{cccc}
SU(2)_{_{(1)} }&0&0&0
\\0&SU(2)_{_{(1')} }&0&0
\\0&0&SU(2)_{_{(2)} }&0
\\0&0&0&SU(2)_{_{(2')} }\ea\right)\,.\ee
As discussed in \cite{gt1}, $\hat{\mf{f}}$ admits also an 
 orthogonal $\mathbb{Z}_2$ decomposition 
 \def \rmf {{\rm f}}
\be\la{uyu}
\hat{\mf{f}}=\hat{\mf{f}}^\perp\oplus\hat{\mf{f}}^\parallel\ , \ \ \ \ \ \ 
\com{\hat{\mf{f}}^\perp}{\hat{\mf{f}}^\perp}\subset\hat{\mf{f}}^\perp\,,\ \ \  \com{\hat{\mf{f}}^\parallel}{\hat{\mf{f}}^\parallel}\subset\hat{\mf{f}}^\perp\,, \ \ \ \com{\hat{\mf{f}}^\perp}{\hat{\mf{f}}^\parallel}\subset\hat{\mf{f}}^\parallel\,,
\ee
defined  by $T$  as follows ($\rmf  \in \hat{\mf{f}}$)
\be 
\rmf=\rmf^\perp+\rmf^\parallel\ , \ \ \ \ \ \ \
[T, \rmf^\perp ]=0\ , \ \ \   \ \ \ \{T, \rmf^\parallel\}=0\ ,
\ee
i.e. $\com{T}{f^\parallel}=2Tf^\parallel$,\  $\trm{STr}(f^\perp f^\parallel)=0$, etc. 
Each of the subspaces in \rf{fef} then splits  into $\perp$ and $\parallel$  parts 
and we can  make the following identifications of the bosonic subspaces 
with the ones in \rf{aa},\rf{aac}
\be
 \mf{f}^\perp_0=\mf{h}= \su(2) \oplus \su(2) \oplus \su(2) \oplus \su(2) \ ,\ \ \ \ \ \  \mf{f}^\parallel_0=\mf{m} \ ,  \ \ \ \ \ \ 
 \mf{f}^\perp_2= \mf{a}= \{ T \} \ , \ \ \ \  \mf{f}^\parallel_2=\mf{n}\ .
\ee
The particular  bases for $\mf{f}_0^\parallel,\,\mf{f}_1^\parallel,\,\mf{f}_3^\parallel$ that we use are given in Appendix \ref{B}.

\subsection{Fermionic part  of the action}

The  action for the Pohlmeyer reduction of the 
$AdS_5 \x S^5$ superstring sigma model is given by  
\rf{act} where \be 
g \in G=Sp(2,2) \times Sp(4)\ , \ \ \ \ \ \
A_\pm \in \mf{f}^\perp_0= \su(2) \oplus \su(2)
 \oplus \su(2) \oplus \su(2) \ ,\ \ \ \ 
 \psr\in\mf{f}_1^\parallel\,,\hs{10pt}\psl\in\mf{f}_3^\parallel\,\,.\no
\ee
These fields  originate from particular  components of the $\psu$ 
currents  \ci{gt1}.\foot{The coset parts of the currents 
that solve the Virasoro  condition are $P_+=\mu T, \ \  P_-=\mu g^{-1} T g$;
$A_\pm$ are  $\mf{h}$ parts of the currents (which are subject to 
the Maurer-Cartan condition in  the first-order formulation).
The fermionic fields are equal (up to $\fr{1}{ \sqrt \mu}$  factor) 
 to the components of the fermionic currents subject to a particular 
 $\kappa$-symmetry  gauge.} 

The  expansion of the bosonic part of the action, i.e. \eqref{bact}, 
 was already   discussed 
in  section 2  and here we turn to the fermionic part
given by
\be\la{fact}
{S}_F=k \;\int d^2x \; \mathcal{L}_F \ , \ \ \ \ \ \
\ \ \ \  \mathcal{L}_F =  \Str \Big(\psl T D_+ \psl+\psr T D_-\psr
+\mu g^{-1}\psl g\psr\Big) \ . 
\ee
 The full action \eqref{act} is  invariant under the $H$-gauge transformations 
 \rf{gag}   with  $\psr$ and $\psl$ transforming as  
\be\ba{cc}
\psr\ra h^{-1}\psr h\,, \ \ \ \ \ \ \ \ \ \ \ \   \psl\ra h^{-1}\psl h\,. \la{trap}
\ea\ee
To expand \rf{fact} near $g=\id$ we use   again \rf{lf1},\rf{decomp}, i.e. 
\be g=e^\eta,\,\hs{20pt}\eta=\M+\xi \ \in \mf{f}_0 \,, \ \ \ \ \ \ 
\M \in \mf{f}_0^\parallel\,,\hs{10pt}\xi \in \mf{f}_0^\perp\,.
\ee
Then (cf. \rf{lagex})
\be
\mathcal{L}_F=\trm{STr}\Big(\psl T \dpl \psl+\psr T \dm\psr+\psl T \com{A_+}{ \psl}+\psr T \com{A_-}{\psr}+\mu\;\psl e^{\mathfrak{D}_\eta}(\psr)\Big)\,.
\ee
As our  aim is to  compute the two-particle   S-matrix 
for the physical fields $(\M, \psl,\psr)$
we  need to expand  the action to quartic order in the  fields only 
 ($
\mathcal{L}_F= \sum_n \mathcal{L}^{(n)}_F$)
\be\la{fqcq}\ba{c}
\mathcal{L}^{(2)}_F=\trm{STr}\Big(\psl T \dpl \psl+\psr T \dm\psr+\mu\;\psl\psr\Big)\,,
\\\mathcal{L}^{(3)}_F=\trm{STr}\Big(A_+\com{ \psl}{\psl T}+ A_-\com{\psr}{\psr T}+\mu\;\xi\com{\psr}{\psl}\Big)\,,
\\\mathcal{L}^{(4)}_F=\trm{STr}\Big(\frac{\mu}{2}\;\psl\com{\M}{\com{\M}{\psr}}+\frac{\mu}{2}\;\psl\com{\xi}{\com{\xi}{\psr}}\Big)\,.
\ea\ee
As follows from  the quadratic Lagrangian,   the
 fermions  $\psr,\,\psl$ have the same mass $\mu$ as the bosons $\M$ (cf. \rf{eoml}):
 their linearized equations of motion are 
 \be\la{eom}
\dpl \psl=2\mu T \psr\,,\hs{20pt}\dm \psr=2\mu T \psl\,, \ \ \ \ \ \ \ 
\ \dpl \dm \psi_{_{L,R}} +\mu^2 \psi_{_{L,R}} =0\,.
\ee

\subsection{Gauge fixing and equivalent local Lagrangian\la{index}}

Following the discussion in section 
 \ref{smb} we will choose the gauge  $A_+=0$, now in the complete action \rf{act}
 including the fermions. 
  We can then integrate out  $A_-$
  getting a constraint  on $\xi$ which generalizes \rf{cons} 
  \be
-\dpl \xi + \fr{1}{2}\com{\M}{\dpl \M}+\com{\psr}{\psr T}+ \fr{1}{2}\com{\xi}{\dpl \xi}+ ...=0\,.
\ee
This constraint  allows  us to eliminate $\xi$  from the  action using  (cf. \rf{xi})
\be\label{xi2}
\xi=\frac{1}{2}\fr{1}{\dpl}\com{\M}{\dpl \M}+\fr{1}{\dpl}\com{\psr}{\psr T}+...\,.
\ee
Dots here stand for higher order terms we will not need. 
From the expansion of the bosonic  \eqref{quadlag},\eqref{cub},\eqref{quart}
 and the fermionic  \eqref{fqcq} parts of the action  we then get 
  the gauge-fixed Lagrangian for $\M,\psl,\psr$
  that generalizes \rf{l1} and again contains no cubic interaction terms, i.e.  
  $\mathcal{L}= \mathcal{L}_4 + 
  \mathcal{O}(\M^5, \M^3 \psi^2,...)$, where 
\be\la{fulllagp}
\begin{split}
\mathcal{L}_4
\,=&\,\trm{STr}\Big(\hs{3pt}\frac{1}{2}\dpl \M\dm \M-\frac{\mu^2}{2}\M^2+\psl T\dpl\psl+\psr T\dm \psr+\mu\psl\psr
\\&\, \quad\hs{5pt}+\fr{1}{8}\com{\M}{\dpl \M}\frac{\dm}{\dpl}\com{\M}{\dpl \M}+\fr{1}{4}\com{\dm \M}{\dpl \M}\fr{1}{\dpl}\com{\M}{\dpl \M}
\\& \,\quad\hs{5pt}-\fr{1}{24}\com{\M}{\dpl \M}\com{\M}{\dm \M}+\fr{\mu^2}{24}\com{\M}{\com{\M}{T}}^2
\\&\, \quad\hs{5pt}+\fr{1}{4}\com{\M}{\dpl \M}\fr{\dm}{\dpl}\com{\psr}{\psr T}+\fr{1}{4}\com{\psr}{\psr T}\fr{\dm}{\dpl}\com{\M}{\dpl \M}
\\&\, \quad\hs{5pt}+\fr{1}{2}\com{\dm \M}{\dpl \M}\fr{1}{\dpl}\com{\psr}{\psr T}+\fr{\mu}{2}\com{\psr}{\psl}\fr{1}{\dpl}\com{\M}{\dpl \M}-\fr{\mu}{2}\com{\M}{\psl}\com{\M}{\psr}
\\& \,\quad\hs{5pt}
+\fr{1}{2}\com{\psr}{\psr T}\fr{\dm}{\dpl}\com{\psr}{\psr T}+\mu \com{\psr}{\psl}\fr{1}{\dpl}\com{\psr}{\psr T}\Big)\,.
\end{split}
\ee
As in \rf{l1} we can  use integration by parts to 
put the bosonic part of   \rf{fulllagp}  into the manifestly local form  $\eqref{l2}$.
The quartic  terms containing fermions
can  also be simplified with the help of  integration by parts 
and  field redefinitions that amount to use of the 
 linearized equations of motion 
$\del_+ \del_- \M + \mu^2 \M=0$ and \rf{eom} in the quartic terms in 
\rf{fulllagp} (see Appendix A).\foot{For  example,  if we 
have a quartic term  containing $\com{\M}{\dpl\dm \M}$ 
we can ignore it: if $\M$ satisfies the linearised equations of motion then 
$\dpl\dm \M \propto \M$ and thus such term vanishes. Equivalently, such term can be removed by a
field redefinition.} 
 The result is an equivalent  local    Lagrangian \rf{spl}  which leads 
to the same  two-particle S-matrix
\be\la{fulllag}
\begin{split}
\mathcal{L}'_4\,=&\,\trm{STr}\Big(\hs{3pt}\frac{1}{2}\dpl \M\dm \M-\frac{\mu^2}{2}\M^2+\psl T\dpl\psl+\psr T\dm \psr+\mu\psl\psr
\\&\, \quad\hs{5pt}+\fr{1}{12}\com{\M}{\dpl \M}\com{\M}{\dm \M}+\frac{\mu^2}{24}\com{\M}{\com{\M}{T}}^2
\\&\, \quad\hs{5pt}-\fr{1}{4}\com{ \psl T}{\psl}\com{\M}{\dpl \M}-\fr{1}{4}\com{\psr}{ T \psr}\com{\M}{\dm \M}-\frac{\mu}{2}\com{\M}{\psr}\com{\M}{\psl}
\\&\, \quad\hs{5pt} +\fr{1}{2}\com{ \psl T}{\psl}\com{\psr}{ T \psr}\Big)\,.
\end{split}
\ee
We expect that this procedure can be continued to all 
orders in the fields, i.e. the all-order generalization of \rf{fulllagp} 
can be transformed into a local  Lagrangian for $\M,\psl,\psr$ 
having   the same  S-matrix as 
  by the equivalence theorem \cite{smateq}.\foot{As in the purely bosonic case, 
 this expectation is supported  by the 
 fact that in the gauge where the local $H$ symmetry is fixed on $g$, the integration over   $A_\pm$  produces a  local   Lagrangian for the coset degrees of freedom and the fermions \ci{gt1}.}

Like the bosonic Lagrangian \rf{l2}, the Lagrangian \rf{fulllagp} or \rf{fulllag} 
is invariant under the residual  global $H=SU(2)^4 $ symmetry transformations 
(cf. \rf{he},\rf{ssss}) 
\be \la{hee}
\M \to  h^{-1} \M h  \ , \ \ \ \ \ \   
\psl \to  h^{-1} \psl h  \ ,  \ \ \ \ \psr \to  h^{-1} \psr h  \ ,\ \ \ \ \
  h=\const  \in H    \  ,  \ee
and so that the resulting  S-matrix should then 
have at least $SU(2)^4$ global   symmetry.

\


Let us now write the Lagrangian  \rf{fulllag} in component notation 
as in \rf{lmgwzw}.
We shall 
 decompose $\M$ and $\psr,\,\psl$ into parts which transform 
 in the bi-fundamental representations of pairs   of the four  $SU(2)$ groups
 that form $H$,
\be\ba{c}
\M=Y+Z\,, \ \ \ \ \ \ \ \ 
\psr=\zer+\chr\,,\hs{15pt}\psl=\zel+\chl\,,
\ea\ee
where  the fields $Y,Z,\zer,\chr,\zel,\chl$
 may be identified as  $2\by 2$ blocks 
of the  $8\x8$ matrix of  $\psu$, see \cite{rtfin} for details.
Schematically, we then have  for  a corresponding 
matrix in $\psu$ (cf. \rf{ssss})
\be\la{diagf}
\left(\ba{cccc}
\su(2)_{_{(1)}}&Y&0&\zeta
\\Y&\su(2)_{_{(1')}}&\chi&0
\\0&\chi&\su(2)_{_{(2)}}&Z
\\\zeta&0&Z&\su(2)_{_{(2')}}\ea\right)\,.\ee
The explicit bases of the relevant subspaces of $\psu$ are given in Appendix \ref{B}.
$Y$ and $Z$  each  represent  $4$ bosonic degrees of freedom of the reduced theory 
for  $AdS_5$ and $S^5$ parts  respectively, while the  $4$ components of
$\zeta_{_{L,R}}$ and $4$ components of $\chi_{_{L,R}}$ are the 
fermions that ``intertwine'' them.

One way to  write  
the component Lagrangian in a simple form 
 is to  formally identify 
the actions  of the pairs of $SU(2)$ 
subgroups of $H$  
as   $SU(2)_{_{(1)}}=SU(2)_{_{(2)}}$ and $SU(2)_{_{(1')}}=SU(2)_{_{(2')}}$  in \rf{ssss}
leaving  a  single ``diagonal''
 $SU(2)\times SU(2)$ subgroup of $H$.
Then  \eqref{diagf}  implies that all the physical fields  will 
be in the  same bi-fundamental representation of this 
 $SU(2)\times SU(2)$  which 
 is locally isomorphic to $SO(4)$, with the 
  bi-fundamental representation of the former being the same as the 
   vector representation of the latter. 
  In this  way  we can rewrite the theory 
  so that  all the fields ($Y,\,Z,\,\zer,\,\chr,\,\zel,\,\chl$) 
  are labelled  by indices of  the same vector representation of this
   $SO(4)$  group. 

Using the basis of the $\psu$  defined  in Appendix \ref{B}, we may 
introduce the component fields as follows 
\be\la{dfact}
\ba{cc}
Y=Y^m T^A_m\,, \,\,& Z=Z^m T^S_m\,, \,\,
\\\zer=e^{\fr{i\pi}{4}}\zer^m T^{R_1}_m\,, \,\,&\chr=e^{\fr{i\pi}{4}}\chr^m T^{R_2}_m\,, \,\,
\\\zel=e^{\fr{i\pi}{4}}\zel^m T^{L_1}_m\,, \,\,&\chl=e^{\fr{i\pi}{4}}\chl^m T^{L_2}_m\,. \,\, 
\ea
\ee
We will use  $m,n, ...=1,2,3,4$ for   $SO(4)$ vector indices
 which will be  raised and lowered with $\delta_{mn}$. 
 $Y^m$ and $Z^m$ are real bosonic  fields and $\zer^m$, $\zel^m$, $\chr^m$ and $\chl^m$ are real (hermitian)  Grassmann  fields.
Making use of the  identities in Appendix \ref{B} we can then 
rewrite \eqref{fulllag} as 
\bea\nn
\mathcal{L}'_4\,&=&\,\frac{1}{2}\dpl Y^m \dm Y_m-\frac{\mu^2}{2}Y^mY_m+\frac{1}{2}\dpl Z^m \dm Z_m-\frac{\mu^2}{2}Z^mZ_m
\\&&\nn+\frac{i}{2}\zel^m\dpl \zeta_{_{L}m}+\frac{i}{2}\zer^m\dm \zeta_{_{R}m}+\frac{i}{2}\chl^m\dpl \chi_{_{L}m}+\frac{i}{2}\chr^m\dm \chi_{_{R}m}-i\mu\zel^m\zeta_{_{R}m}-i\mu\chl^m\chi_{_{R}m}
\\&&\nn-\fr{1}{12}\left(Y^mY^p\dpl Y^n  \dm Y^q-Z^mZ^p\dpl Z^n  \dm Z^q\right)\left(\delta_{mp}\delta_{nq}-\delta_{mq}\delta_{np}\right)
\\&&-\frac{\mu^2}{24}(Y^mY_mY^nY_n-Z^mZ_mZ^nZ_n) \la{theone}
\\&&\nn-\fr{i}{16} C_{mnpq} \big(\zel^m\zel^nY^p\dpl Y^q+\zer^m\zer^n Y^p\dm Y^q   +  \zel^m\zel^qZ^n \dpl Z^p+\zer^m\zer^q  Z^n\dm Z^p 
\\&&\nn   +   \chl^m\chl^qY^p\dpl Y^n+\chr^m\chr^q Y^p\dm Y^n
+   \chl^m\chl^nZ^q\dpl Z^p+\chr^m\chr^n Z^q\dm Z^p\big)
\\&&\nn+\fr{i\mu}{8}( \zer^m\zeta_{_{L}m} + \chr^m\chi_{_{L}m})
 (   Y^nY_n- Z^nZ_n) +\fr{i\mu}{4}C_{mnpq} ( \zer^m\chl^pY^qZ^n -\chr^m\zel^pY^nZ^q)  
\\&&\nn+\fr{1}{8}\epsilon_{mnpq} ( \zel^m\zel^n\zer^p\zer^q- \chl^m\chl^n\chr^p\chr^q)\,,
\eea
where
\be
C_{mnpq}\equiv - \epsilon_{mnpq}-\delta_{mp}\delta_{nq}+\delta_{mq}\delta_{np}+\delta_{mn}\delta_{pq}\,.
\ee
This  component form of the  Lagrangian \rf{fulllag}  has the 
advantage of  simplicity  but it  has 
 only $SO(4)$ or $SU(2) \times SU(2)$ part of global symmetry $H$ as its manifest symmetry -- 
 another $SU(2)\times SU(2)$ is hidden.
 
 
It   is  possible  to write  \rf{fulllag} also 
 in a manifestly $SU(2)^4$ invariant form  by 
   trading  each $SO(4)$ index for a pair of $SU(2)$ ones so 
   as to make it explicit that the fields  belong to bi-fundamental
    representations of the four  $SU(2)$ groups according to \eqref{diagf}. 
    To do so  let us  introduce  the indices $a,\,\adot,\,\al,\ald$ 
corresponding to the 
 fundamental representations of the four $SU(2)$'s in \rf{ssss}. 
 Then  we may re-label the  fields as follows\foot{The 2-indices are raised and lowered with 
 the antisymmetric tensors $\epsilon^{ab}$, etc., i.e. $F^a = \epsilon^{ab}  F_b, \ 
 F_b= \epsilon_{bc} F^c$. 
 Dotted and undotted indices are  assumed to be completely independent. 
 We use the convention that $\epsilon^{12}=1,\,\epsilon_{12}=-1$,  $\epsilon^{ab}
 \epsilon_{bc}=\delta^a_c$
 and the rescaled  set of Pauli matrices 
 $$\si^1=\bars^1=\fr{1}{\sqrt{2}}\bp1&0\\ 0&1\emp \ ,  
 \ \ \ \   \si^2=-\bars^2=\fr{1}{\sqrt{2}}\bp  0&1\\-1&0\emp \,, \ \  \ \  
 \si^3=-\bars^3=\fr{1}{\sqrt{2}}\bp 0&i\\i&0\emp \,, \ \ \ \ 
  \si^4=-\bars^4=\fr{1}{\sqrt{2}}\bp i&0\\0&-i\emp \,.
 $$
 } 
\be\la{trans}\ba{cc}
Y^m=(\bars^m)^{\adot a}Y_{a\adot}\,,\;\;&Y_{a\adot}=(\si_m)_{a \adot}Y^m\,,\;\;
\\Z^m=(\bars^m)^{\ald \al }Z_{\al \ald}\,,\;\;&Z_{\al \ald}=(\si_m)_{\al \ald}Z^m\,,\;\;
\\\zeta^m=(\bars^m)^{\ald a}\zeta_{a\ald}\,,\;\;&\zeta_{a\ald}=(\si_m)_{a \ald}\zeta^m\,,\;\;
\\\chi^m=(\bars^m)^{\adot \al}\chi_{\al\adot}\,,\;\;&\chi_{\al\adot}=(\si_m)_{\al \adot}\chi^m\,
\;\;
\ea\ee
where  $(Y_{a\adot})^*= Y^{\adot a }$, etc. 
 The explicit  result  of the translation of \rf{theone} into this 
 manifestly  $SU(2)^4$  invariant  form is
  \bea
\mathcal{L}'_4\;=&&\ha\dpl Y_{a\adot}\dm Y^{\adot a}-\fr{\mu^2}{2}Y_{a\adot}Y^{\adot a}+\ha\dpl Z_{\al\ald}
\dm Z^{\ald\al}-\fr{\mu^2}{2}Z_{\al\ald} Z^{\ald\al}\nn
\no \\ &&+\fr{i}{2}{\zel}_{a\ald}\dpl\zel^{\ald a}+\fr{i}{2}{\zer}_{a\ald}\dm\zer^{\ald a}+\fr{i}{2}{\chl}_{\al
\adot}\dpl\chl^{\adot \al}+\fr{i}{2}{\chr}_{\al\adot}\dm\chr^{\adot \al}-i\mu{\zel}_{a\ald}\zer^{\ald a}-i\mu{
\chl}_{\al\adot}\chr^{\adot\al}\nn
\no \\ &&-\fr{1}{12}\big(Y_{a\adot}Y^{\adot a}\dpl Y_{b\bdot}\dm Y^{\bdot b}-Y_{a\adot}\dpl Y^{\adot a} Y_{b 
\bdot} \dm Y^{\bdot b}+\fr{\mu^2}{2}Y_{a\adot}Y^{\adot a}Y_{b\bdot}Y^{\bdot b}\big)\nn
\no \\ &&+\fr{1}{12}\big(Z_{\al\ald}Z^{\ald \al}\dpl Z_{\bet\betd}\dm Z^{\betd \bet}-Z_{\al\ald}\dpl Z^{\ald 
\al} Z_{\bet \betd} \dm Z^{\betd \bet}+\fr{\mu^2}{2}Z_{\al\ald}Z^{\ald \al}Z_{\bet\betd}Z^{\betd \bet}\big)\nn
\no \\ &&+\fr{i}{8}\big({\zel}_{a\ald}{\zel}^{\ald b}Y^{\bdot a}\dpl Y_{b\bdot}+{\zer}_{a\ald}{\zer}^{\ald b}Y^{\bdot 
a}\dm Y_{b\bdot}+\mu\,{\zer}_{a\ald}\zel^{\ald a}Y_{b\bdot}Y^{\bdot b}\big)\la{sut}
 \\ &&-\fr{i}{8}\big({\zel}_{a\ald}{\zel}^{\betd a}Z^{\ald \bet}\dpl Z_{\bet \betd}+{\zer}_{a\ald}{\zer}^{\betd 
a}Z^{\ald \bet}\dm Z_{\bet \betd}+\mu\,{\zer}_{a\ald}\zel^{\ald a}Z_{\bet\betd}Z^{\betd \bet}\big)\nn
\no \\ &&+\fr{i}{8}\big({\chl}_{\al\adot}{\chl}^{\bdot \al}Y^{\adot b}\dpl Y_{b\bdot}+{\chr}_{\al\adot}{\chr}^{\bdot
 \al}Y^{\adot b}\dm Y_{b\bdot}+\mu\,{\chr}_{\al\adot}\chl^{\adot \al}Y_{b\bdot}Y^{\bdot b}\big)\nn
\no \\ &&-\fr{i}{8}\big({\chl}_{\al\adot}{\chl}^{\adot \bet}Z^{\betd \al}\dpl Z_{\bet\betd}+{\chr}_{\al\adot}{\chr}^{\adot 
\bet}Z^{\betd \al}\dm Z_{\bet\betd}+\mu\,{\chr}_{\al\adot}\chl^{\adot \al}Z_{\bet\betd}Z^{\betd \bet}\big)\nn
\cr &&+\fr{i\mu}{2}\big({\zer}_{a\ald}{\chl}_{\bet\bdot}Y^{\bdot a}Z^{\ald \bet}
-{\chr}_{\al\adot}{\zel}_{b\betd}Y^{\adot b}Z^{\betd \al}\big)+\fr{1}{2}
\big({\zel}_{a\ald}{\zel}_{b\betd}\zer^{\ald b}\zer^{\betd a}
-{\chl}_{\al\adot}{\chl}_{\bet\bdot}\chr^{\adot\bet}\chr^{\bdot\al}\big)\,.
\eea
The Lagrangian \eqref{theone} or \rf{sut}   is also  manifestly 
 2-d Lorentz invariant,  assuming that 
the fermions transform as Majorana-Weyl  2-d  spinors  \ci{gt1}. 

The original reduced  model \ci{rtfin} was shown to be UV 
finite  to the two-loop order (and conjectured to be finite to all orders) 
in \ci{rtfin};  the same should of course be true also 
for its gauge-fixed   version  \rf{fulllag} or \rf{theone}.
This implies that the  corresponding quantum S-matrix should  be 
finite, i.e.  should depend  only on the  original tree-level mass  scale $\mu$.

\subsection{Special cases: reduced Lagrangians for $AdS_2\x S^2$ and $AdS_3\x S^3$}

Let us  now look at some special cases  of \eqref{theone}
corresponding to the  reductions of the superstring theory  on $AdS_2\x S^2$
 \cite{gt1}
 and on $AdS_3 \x S^3$  \cite{gt2}.

Replacing $PSU(2,2|4)$ with $PSU(1,1|2)$ gives superstring theory on $AdS_2 \x S^2$
and the corresponding reduced  theory 
 can be identified \ci{gt1}  with the (2,2)  supersymmetric sine-Gordon model \cite{susyn2}.
  Its  Lagrangian may be written as \cite{gt1}\foot{Compared to \cite{gt1}
   we have redefined $\alpha\ra-e^{\fr{i\pi}{4}}\alpha$, $\delta\ra e^{\fr{i\pi}{4}}\delta$, 
   $\nu\ra e^{\fr{i\pi}{4}}\nu$, $\rho\ra e^{\fr{i\pi}{4}}\rho$ in order  to 
 get  the  standard  hermitian 
 conjugation property $(ab)^*=b^*a^*$ for the Grassmann fields, as opposed to 
 the convention $(ab)^*=a^*b^*=- b^*a^*$
 used in \cite{gt1}.
 This is also the origin of the $e^{\fr{i\pi}{4}}$ factors in \rf{dfact}. 
 Note also that in  \cite{gt1}
  the Lagrangian was  rescaled by a factor of $\fr{1}{2}$.
  } 
\bea\nn
\mathcal{L}&=&
2\Big[
\dpl\phi\dm\phi+\dpl\vp\dm\vp+\fr{\mu^2}{2}\left(\cos 2\vp-\cosh 2\phi \right)+i\,\alpha\dm\alpha+i\,\delta\dm\delta+i\,\nu\dpl\nu+i\,\rho\dpl\rho
\\&&\la{susy2}\hs{15pt}-2i\mu\Big(\cosh\phi\cos\vp\ \left(\nu\delta+\rho\alpha\right)+\sinh\phi\sin\vp\ (-\rho\delta+\nu\alpha)\Big)
\Big]
\,.
\eea
Here $\phi,\,\vp$ are real bosonic  fields and $\alpha,\,\delta,\,\nu,\,\rho$ are real (hermitian) fermions.
Expanding this  Lagrangian  to quartic order, rescaling the fields by $\fr{1}{2}$  and 
renaming  them,  
one finds that it  becomes  the same as \rf{theone} if   all $SO(4)$  indices there 
take just only  one value, i.e. 
\be\la{s2}
\begin{split}
\mathcal{L}_4=&\frac{1}{2}\dpl Y \dm Y-\frac{\mu^2}{2}Y^2+\frac{1}{2}\dpl Z \dm Z-\frac{\mu^2}{2}Z^2\\&+\frac{i}{2}\zel\dpl \zeta_{_{L}}+\frac{i}{2}\zer\dm \zeta_{_{R}}+\frac{i}{2}\chl\dpl \chi_{_{L}}+\frac{i}{2}\chr\dm \chi_{_{R}}-i\mu\zel\zeta_{_{R}}-i\mu\chl\chi_{_{R}}
\\&-\frac{\mu^2}{24}(Y^4-Z^4) +\fr{i\mu}{8}( \zer\zel + \chr\chl ) (Y^2- Z^2) +
\fr{i\mu}{4}(\zer\chl - \chr\zel) YZ \,.
\end{split}\ee
Like the action for \rf{susy2}, the   action for   \rf{s2} is thus invariant 
 (2,2)  supersymmetry. As discussed in \ci{gt1,gt2,rtfin} it is  an open  question  
 if the general action \rf{act}  and thus \rf{theone}   may also be invariant under 
  (a properly defined) 2-d supersymmetry  (cf. also section 5).

\

The  action  that arises from the Pohlmeyer reduction of the $AdS_3\x S^3$
superstring theory with asymmetric (axial)   gauging of $H=U(1)\x U(1)$
 can be found, e.g., 
 by  fixing  a gauge on $g$  and 
  integrating  out $A_\pm$. 
   This  gives  a local Lagrangian (with regular expansion near 
  the  trivial  vacuum)  which is 
   a fermionic extension of the sum of the 
   complex sine-Gordon  and the complex sinh-Gordon models 
   (see \ci{gt2}   and section \ref{tauaut}
  for details)\foot{As in the $AdS_2 \x S^2$ case discussed above 
  we  have redefined $\alpha \ra-\alpha$, $\beta\ra-\beta$ and 
  rescaled all the fermions by $e^{\fr{i\pi}{4}}$ compared
   to \cite{gt2}. Again,  an extra  overall factor in the Lagrangian  arises as we 
   have not rescaled it  as was done in \cite{gt2}.}
\bea
\nn\mathcal{L}\!\!&=&\!\!\ 2\bigg[
\dpl\phi\dm\phi+\tanh^2\phi\ \dpl v \dm  v +\dpl\vp\dm\vp+\tan^2\vp\ \dpl u\dm u+\fr{\mu^2}{2}
\left(\cos 2\vp-\cosh 2\phi\right)
 \\\nn&&+i\,\alpha\dm\alpha+i\,\beta\dm\beta+i\,\gamma\dm\gamma+i\,\delta\dm\delta+
i\,\lambda\dpl\lambda+i\,\nu\dpl\nu+i\,\rho\dpl\rho+i\,\sigma\dpl\sigma \la{2} 
\\&&-i\tanh^2\phi\big[\dpl v\left(\lambda\nu-\rho\sigma\right)-\dm v\left(\alpha\beta-\gamma\delta\right)\big]
\\\nn&&+i\tan^2\vp\big[\dpl u\left(\lambda\nu-\rho\sigma\right)-\dm  u\left(\alpha\beta-\gamma\delta\right)\big] 
\\\nn&&+ ({\sec^2\vp}-{\sech^2\phi}) 
\left(\alpha\beta-\gamma\delta\right)\left(\lambda\nu-\rho\sigma\right)-2i\mu\Big(\cosh\phi\cos\vp\left(\lambda\gamma+\nu\delta+\rho\alpha+\sigma\beta\right)
\\\nn&&+\sinh\phi\sin\vp\big[\cos(v+ u)(\rho\delta-\sigma\gamma+
\lambda\beta-\nu\alpha)-\sin(v+ u)(\lambda\alpha+\nu\beta-\rho\gamma-\sigma\delta)\big]\Big)
\bigg]
\,.
\eea
Here $\phi,\,\vp,\,v,\, u$ are real commuting  and $\alpha,\,\beta,\,\gamma,\,
\delta,\,\lambda,\,\nu,\,\rho,\,\sigma$ are real anticommuting fields. 
Expanding  this to quartic order in ``radial'' directions $\phi,\vp$ 
as in  section \ref{tauaut}  
and using the following field  redefinition 
\bea 
&& Y_1= 2  \phi\  \cosh v \ , \ \ \ \ Y_2= 2  \phi\  \sinh v \ , \ \ \ \ 
Z_1= 2  \vp\  \cos u \ , \ \ \ \ Z_2= 2 \vp\  \sin u \ ,  \no\\
&&( \zer^1,\ \zer^2,\ \zel^1,\ \zel^2, \ \chr^1,\ \chr^2,\  \chl^1,\ 
\chl^2) = 
2 (\alpha,\beta,\rho,\sigma,\delta, \gamma, \nu, \lambda)
\,, \la{redf}
\eea
we  conclude that the resulting Lagrangian 
becomes the same as \rf{theone}  with $m,n,p,q=1,2$ (e.g., the terms containing $\epsilon_{mnpq}$
 in \rf{theone} drop out). 
 
 

\renewcommand{\theequation}{4.\arabic{equation}}
\setcounter{equation}{0}

\section{Tree-level S-matrix of   reduced  theory \\
for   $AdS_5 \x S^5$ superstring \la{tlta}}

Starting with the component Lagrangian \rf{theone} it is straightforward to read-off  the corresponding  tree-level two-particle S-matrix
following the same steps as in section \rf{tl1}. Again, 
 we rescale the fields by $\fr{1}{\sqrt{k}}$ and carry out the expansion of the S-matrix, 
\rf{expansion}, in powers of $1/k$. The quadratic part of \rf{theone} describes  4+4 bosonic and 4+4 fermionic massive   degrees of freedom for which we  have  the following mode expansion (cf. \rf{md})
\be\ba{c}\la{yd}
Y^m(x)=\int\fr{dp}{2\pi}\;\fr{1}{\sqrt{2 \epsilon}}\left(a_{Yp}^m e^{-i \Vec{p}\cdot \Vec{x}}+a_{Yp}^{m\dagger} e^{i \Vec{p}\cdot \Vec{x}}\right)\Big|_{\epsilon\,=\,\epsilon_{p}}\,,
\\\zel^m(x)=\int\fr{dp}{2\pi}\;\fr{1}{\sqrt{2 \epsilon}}\left(u_{\zel}(p)\ a_{\zeta p}^m e^{-i \Vec{p}\cdot \Vec{x}}+v_{\zel}(p)\ a_{\zeta p}^{m\dagger} e^{i \Vec{p}\cdot \Vec{x}}\right)\Big|_{\epsilon\,=\,\epsilon_{p}}\,,
\\\zer^m(x)=\int\fr{dp}{2\pi}\;\fr{1}{\sqrt{2 \epsilon}}\left(u_{\zer}(p)\ a_{\zeta p}^m e^{-i \Vec{p}\cdot \Vec{x}}+v_{\zer}(p)\ a_{\zeta p}^{m\dagger} e^{i \Vec{p}\cdot \Vec{x}}\right)\Big|_{\epsilon\,=\,\epsilon_{p}}\,,
\ea\ee
where the fermionic wave functions have  the following explicit form 
in terms of  the rapidity defined in  \eqref{rap} 
\be\la{uuu}
u_{\zel}(p)=\sqrt{\mu}\,i\,e^{-\fr{\theta}{2}}\,, \ \ \ \ 
v_{\zel}(p)=-\sqrt{\mu}\,i\,e^{-\fr{\theta}{2}}\,,
\ \ \ \ \ u_{\zer}(p)=\sqrt{\mu}\,e^{\fr{\theta}{2}}\,, \ \ \ \ \ 
v_{\zer} (p)=\sqrt{\mu}\,e^{\fr{\theta}{2}}\,, 
\ee
and 
\be 
\com{a^{m}_{Yp}}{a^{n\dagger}_{Yp'}}=2\pi \delta^{mn}\delta(p-p')\,,\ \ \ \ 
\ \ \ \ \ \ \ \acom{a^{m}_{\zeta p}}{a^{n\dagger}_{\zeta p'}}=2\pi \delta^{mn}\delta(p-p')\,. 
\ee 
There are also similar relations for $Z^m,\,\chl^m,\,\chr^m$.
Then the  normal ordered quadratic Hamiltonian $\H_2 $ has standard free oscillator form, i.e.   its action on one-particle states is 
\be
\H_2  a^{m\dagger}_{_{\Phi}p}\ket{0}=\epsilon_p\;a^{ m\dagger}_{_{\Phi}p}\ket{0} \ , 
\ee
where $\Phi$ stands for any of the fields  
 $Y,\,Z,\,\zeta,\,\chi$. 


As in the bosonic case  in section \rf{tl1}, we
will  write the S-matrix in terms of the $\tm$-matrix defined in \eqref{expansion}, 
which is again determined by the normal-ordered quartic part  
 of \eqref{fulllag}.
Plugging in the mode decompositions  \eqref{yd}
 we may  compute the action of $\mbb{T}$ on the 
 two-particle initial states 
\be
\ket{\Phi_{1}^m(p_1)\Phi_2^n(p_2)}=2\sqrt{\epsilon_{p_1}\epsilon_{p_2}}a^{m\dagger}_{_{\Phi_1} p_1}a^{n\dagger}_{_{\Phi_2}p_2}\ket{0}\,.
\ee
We shall again use the definition of rapidities in \rf{vtd}. For simplicity we will assume $p_1>p_2 \;\Leftrightarrow \theta_1 >\theta_2$.
 This leads to  the expression  for the 
 $\tm$-matrix as a function of $\vt$ in terms of one type of 
  $SO(4)$ vector indices  which we present 
    explicitly in Appendix \ref{C}.

  To write the $\tm$-matrix in 
  the form exhibiting 
  the full bosonic symmetry group $SU(2)^4$  
   we  shall trade each $SO(4)$ 
   index for a pair of $SU(2)$ ones
    so as to make it explicit that the fields 
     belong to the bi-fundamental representations of the four  
     $SU(2)$ groups according to \eqref{diagf} as described in section \ref{index} (see \rf{trans}).
We present the resulting $SU(2)^4$ invariant form of the $\tm$-matrix
in  Appendix \ref{C}. For example, we get
  \begin{footnotesize}
\bea \nn
\mbb{T}\ket{Y_{a\adot}(p_1)Y_{b\bdot}(p_2)}&=&\fr{1}{4\sinh\vt}\Big[\Big(2\sinh^2\fr{\vt}{2}\;\delta_a^c\delta_b^d\delta_{\adot}^{\cd}\delta_{\bdot}^{\dd}-\cosh\vt \big(\delta_a^c\delta_b^d\delta_{\adot}^{\dd}\delta_{\bdot}^{\cd}+\delta_a^d\delta_b^c\delta_{\adot}^{\cd}\delta_{\bdot}^{\dd}\big)\Big)\ket{Y_{c\cd}(p_1)Y_{d\dd}(p_2)}
\\\nn&&\hs{32pt}-\sinh\fr{\vt}{2}\left(\de_a^c\de_b^d\epsilon_{\adot\bdot}\epsilon^{\gad\ded}
\ket{\zeta_{c\gad}(p_1)\zeta_{d\ded}(p_2)}+\epsilon_{ab}\epsilon^{\ga\de}\de_{\adot}^{
\cd}\de_{\bdot}^{\dd}\ket{\chi_{\ga\cd}(p_1)\chi_{\de\dd}(p_2)}\right)\Big]\,,
\\\nn\mbb{T}\ket{\zeta_{a\ald}(p_1)\zeta_{b\betd}(p_2)}&=&\fr{1}{4\sinh\vt}\Big[-\sinh\fr{  \vt}{2}\left(\de_a^c\de_b^d\epsilon_{\ald\betd}\epsilon^{\cd\dd} \ket{Y_{c\cd}(p_1)Y_{d\dd}(p_2)}-\epsilon_{ab}\epsilon^{\ga\de}\de_{\ald}^{\gad}\de_{\betd}^{\ded}\ket{Z_{\ga\gad}(p_1)Z_{\de\ded}(p_2)}\right) 
\\&&\nn\hs{50pt}+\cosh\vt\left(\de_a^c\de_b^d\de_{\ald}^{\ded}\de_{\betd}^{\gad}-\de_a^d\de_b^c\de_{\ald}^{\gad}\de_{\betd}^{\ded}\right)\ket{\zeta_{c\gad}(p_1)\zeta_{d\ded}(p_2)}\Big]\,,
\\\nn\mbb{T}\ket{Y_{a\adot}(p_1)\zeta_{b\betd}(p_2)}&=&\fr{1}{4\sinh\vt}\Big[\Big(\sinh^2\fr{\vt}{2}\;\de_a^c\de_b^d\de_{\adot}^{\cd}\de_{\betd}^{\ded}-\cosh\vt\de_a^d\de_b^c\de_{\adot}^{\cd}\de_{\betd}^{\ded}\Big)\ket{Y_{c\cd}(p_1)\zeta_{d\ded}(p_2)}
\\\nn&&\hs{32pt}-\cosh\fr{\vt}{2}\;\de_a^c\de_b^d\de_{\adot}^{\dd}\de_{\betd}^{\gad}\ket{\zeta_{c\gad}(p_1)Y_{d\dd}(p_2)}-\sinh\fr{\vt}{2}\;\epsilon_{ab}\epsilon^{\ga\de}\de_{\adot}^{\cd}\de_{\betd}^{\ded}\ket{\chi_{\ga\cd}(p_1)Z_{\de\ded}(p_2)}\Big]\,.
\eea
\end{footnotesize}

For a generic integrable theory with non-simple $G_1 \times G_2$  global symmetry 
and with fields transforming in the bi-fundamental representation of this group 
the S-matrix should exhibit group factorization property 
(see, e.g.,  \ci{orw,review}).\foot{\la{foot2}Let us recall (see \ci{afz,kmrz}) that  this 
 can be understood as a requirement that the 
Faddeev-Zamolodchikov algebra is also a direct
product. Consider the field $F_{a\al}$
 where the index $a$ is from $G_1$ and $\al$ is from $G_2$.
$F_{a\al}$  may  then be represented ``on-shell'' 
by a bilinear term  $u_a v_{\al}$ in oscillators, 
 where  $u_a $
transforms under $G_1 $ and $v_{\al}$ transforms under $G_2$ and the two sets of oscillators 
mutually commute. The braiding
relations for each of these sets are determined by an $G_1$- or $G_2$- invariant S-matrix consistent
with the Lagrangian of the theory.}
Such factorization indeed happens in the light-cone gauge
 $AdS_5 \x S^5$ superstring  S-matrix  which is invariant under the   product supergroup
 $PSU(2|2) \x PSU(2|2) $   \ci{bsmat,af,kmrz,afz,review}.

Since the  reduced  theory is integrable, and  
the fields are in (different) 
bi-fundamental representations of $SU(2) \times SU(2)$
 subgroups of $H=SU(2)^4 $
we  should  expect at least  partial group factorization 
of the two-particle   S-matrix here  as well. 

The field contents of the light-cone superstring and  the reduced theories 
are identical in  how they transform under the bosonic 
symmetry group, $SU(2)^4$.
 Therefore, we  may  expect to find at least  part of the
factorisation of the 
S-matrix
seen  in the 
superstring theory to be present in the reduced theory.

 Remarkably,  it turns out 
that we  get  exactly the same   factorisation structure as in the superstring case, \ci{kmrz}. 
 In general, the S-matrix group factorisation
\be\la{ses}
\mbb{S}=\tilde{\mbb{S}}\otimes \tilde{\mbb{S}}\ ,
\ee
implies the following factorization of the  leading term  in the 
 $\mbb{T}$-matrix (cf. \rf{expansion})\foot{Here $\mbb{I}$ is  the identity operator.}
\be\la{tef} \mbb{T}=\mbb{I}\otimes \tilde{\mbb{T}}+\tilde{\mbb{T}}\otimes \mbb{I}\,,
\ee
 To exhibit the   factorization   let us introduce the super-indices
 \foot{Again, 
   dotted and  undotted indices are completely independent, representing 
   fundamental  representations of the four independent $SU(2)$ groups.}
 $$A=(a|\alpha) \ , \ \ \ \ \  \ \ \   \ \ \ \ \dot{A}=(\adot|\ald)\ , 
 $$
 where the lower-case latin indices are
   Grassmann even and the greek indices are Grassmann odd. 
 We can then describe  all of our  
  fields in terms of one field $\Phi_{A\dot{A}}$. 
  The (centrally extended) $PSU(2|2) \x PSU(2|2)$ factorisation \rf{ses} 
  of
   the superstring  S-matrix   means that \ci{kmrz,review} 
   \be \la{meas}
   {\rm S}_{A\dot{A},B\dot{B}}^{C\dot{C},D\dot{D}}=
   (-1)^{[\dot{A}][B]+[\dot{C}][D]}{\rm S}_{AB}^{CD}
    {\rm S}_{\dot{A}\dot{B}}^{\dot{C}\dot{D}} \ , \ee
    so that  the leading term in the 
    $\mbb{T}$-matrix \rf{tef}  can be written 
   in  the following compact form
\be\bs\la{factor}
\mbb{T}\ket{\Phi_{A\dot{A}}(p_1)\Phi_{B\dot{B}}(p_2)}&=
\fr{1}{4\sinh\vt}\left[(-1)^{[\dot{A}]\left([B]+[D]\right)}
T_{AB}^{CD}\de_{\dot{A}}^{\dot{C}}\de_{\dot{B}}^{\dot{D}}\right.
\\&\left.\hs{50pt}+\ (-1)^{\left([\dot{A}]+[\dot{C}]\right)[D]}
\de_{A}^{C}\de_{B}^{D}T_{\dot{A}\dot{B}}^{\dot{C}\dot{D}}\right]
\ket{\Phi_{C\dot{C}}(p_1)\Phi_{D\dot{D}}(p_2)}
\end{split}\ee
Here $[a]=[\dot{a}]=0$ and $[\alpha]=[\dot{\alpha}]=1$
and  the explicit form of the coefficient 
 $T_{AB}^{CD}$ can be written in terms of ten arbitrary functions $K_i$
 of the kinematic variables (called A,B, ... in \ci{kmrz})
\be\la{kek}
\ba{l}
\;\;\;\;\;\;\;T_{ab}^{cd}=K_1\;\de_a^c\de_b^d+K_2\;\de_a^d\de_b^c\,,
\\\;\;\;\;\;\;\;T_{\al\bet}^{\ga\de}=K_3\;\de_\al^\ga\de_\bet^\de+K_4\;\de_\al^\de\de_\bet^\ga\,,
\\\ba{ll}
\!\!T_{ab}^{\ga\de}=K_5\;\epsilon_{ab}\epsilon^{\ga\de}\,,\;\;\;&T_{\al\bet}^{cd}=K_6\;\epsilon_{\al\bet}\epsilon^{cd}\,,\;\;\;
\\\!\!T_{a\bet}^{\ga d}=K_7\;\de_a^d\de_\bet^\ga\,,&T_{\al b}^{c \de}=K_8\;\de_\al^\de\de_b^c\,,
\\\!\!T_{a\bet}^{c\de}=K_9\;\delta_a^c\delta_\bet^\de\,,&T_{\al b}^{\ga d}=K_{10}\;\delta_\al^\ga\delta_b^d\,.
\ea
\ea
\ee
Remarkably, the   result  for the reduced theory $\mbb{T}$-matrix that we found (and which 
is  presented in detail in Appendix \ref{C})  
 has  exactly   the form    \rf{factor},\rf{kek}  with the functions $K_i$ given by 
\be\la{comp}
\ba{l}
K_1=-K_3=\sinh^2\fr{\vt}{2}
\\K_2=-K_4=-\cosh\vt
\\K_5=K_6=-\sinh\fr{\vt}{2}
\\K_7=K_8=-\cosh\fr{\vt}{2}
\\K_9=-K_{10}=0\,.
\ea
\ee
The Pohlmeyer reduced theory \rf{act}
is 2-d  Lorentz-invariant and therefore the functions $K_i$  depend
only on  the difference of the two  rapidities  $\vt=\theta_1-\theta_2$
 (we have assumed that $\theta_1>\theta_2$ in   \rf{vtd}).

For comparison,   the light-cone superstring $\mbb{T}$-matrix found 
by the explicit computation in  \cite{kmrz}   has  the  form \rf{factor},\rf{kek} 
 with  $K_i$  depending separately on the two rapidities:\foot{The difference
 in sign
  in $K_5$ and $K_6$ between \rf{comp1} and \cite{kmrz} arises from the alternative definitions 
  used for $\epsilon_{ab},\;\epsilon_{\al\bet}$: we  used
  $\epsilon_{12}=-1$, whereas in \cite{kmrz} $\epsilon_{12}=1$.
  In \ci{kmrz} the expansion of the S-matrix was carried out in powers of 
  inverse string tension, $\fr{2\pi}{\sqrt{\lambda}}$
  which played the role of the expansion parameter  $1/k$ in 
  \rf{expansion}.}
 \be\la{comp1}
\ba{l}
K_1=-K_3=(\sinh\theta_1-\sinh\theta_2)^2
\\K_2=-K_4=4\sinh\theta_1\sinh\theta_2
\\K_5=K_6=4\sinh\theta_1\sinh\theta_2\sinh\fr{\theta_1-\theta_2 }{2}
\\K_7=K_8=4\sinh\theta_1\sinh\theta_2\cosh\fr{  \theta_1-\theta_2 }{2}
\\K_9=-K_{10}= -\sinh^2\theta_1+\sinh^2\theta_2 \,.
\ea
\ee 
Note that the vanishing of $K_9$ and $K_{10}$ in \rf{kek},\rf{comp} 
  reflects the fact that the  bosonic part of the reduced theory in \rf{sut}
  is the direct sum of the ``AdS'' and ``sphere'' parts (which separate  as usual in the
  conformal gauge)   while in the light-cone 
  gauge  superstring    action used in \ci{kmrz} the  corresponding sets of the bosonic fields
  were coupled.\foot{Note also that un-extended  $PSU(2|2)$ symmetry would imply, in particular,  that 
  $K_5=K_6=0$  (cf. also \ci{bek}). We thank  N. Beisert and T. McLoughlin 
  for this remark.}


 It is  somewhat unexpected 
  to  find that the reduced theory S-matrix 
  has  the same 
   type of group  factorisation as  
  the superstring theory  S-matrix  in \ci{kmrz} 
  while its    action \rf{act}
  had only the  bosonic $SU(2)^4$  group  as its  manifest  symmetry.\foot{The reduced theory coset 
  \rf{co}  is purely bosonic, so any remaining fermionic supergroup symmetry \rf{coo} 
  of the superstring theory would be ``hidden'' here.} 
   The reduced  theory S-matrix  in \rf{kek},\rf{comp} 
  has the obvious symmetry under interchanging 
  different types of indices $a \to \alpha,$ etc.
  (so that $K_1\to -K_3, \ K_2 \to - K_4, $ etc.)  reminiscent  of a boson-fermion symmetry 
  and exactly the same applied to the 
   superstring case \rf{comp1}.


\renewcommand{\theequation}{5.\arabic{equation}}
\setcounter{equation}{0}

\section{Concluding remarks and open problems}

One of the main conclusions   of the present  paper is that there exists  a  special 
2-d Lorentz  covariant  S-matrix (corresponding  to the local  UV finite  massive 
integrable theory \rf{act}) whose algebraic structure is very similar 
to that of the  S-matrix of the   \adss  superstring theory  in the $S^5$  light-cone gauge. 

An obvious  extension of our tree-level 
S-matrix computation is its analog at the  1-loop level. 
To compute the  1-loop  two-particle S-matrix for the 
elementary fields of the reduced theory  it is again enough to use 
the quartic Lagrangian \rf{theone} or \rf{sut}. The reason is 
 the absence of cubic vertices and 
the  cancellation of   diagrams with tadpoles as the  theory \rf{act} 
is  UV finite \ci{rtfin}. 
Finding again the same group factorization of the 1-loop 
 S-matrix would  be a non-trivial  check of quantum integrability.

Let us now   discuss some   remaining issues and problems. 
An  obvious   question is if 
the two S-matrices (with coefficients in \rf{comp} and \rf{comp1} respectively)
are  related in some way.  
Since  they correspond to integrable  theories  and should  thus
  satisfy (cf. \ci{afz}) 
  the Yang-Baxter equation (which is, in general,  quite constraining)
   it is likely 
 that there is a   transformation mapping one into the other. 

The  group factorization structure of the reduced theory S-matrix 
we  have found suggests that  the reduced theory  should 
possess  a hidden  symmetry  mixing bosons with fermions
like the $PSU(2|2) \x PSU(2|2)$ of the superstring case. 
Since the fermionic   fields  in \rf{act} or \rf{sut} have the standard first-order 
 kinetic terms,  
a  target-space  symmetry relating them to bosons  cannot  be realised 
in a simple local way at the
Lagrangian level.\foot{Note 
   that the    $PSU(2|2) \x PSU(2|2)$
symmetry  on the string side appears 
as a  symmetry  of the  corresponding  light-cone gauge Hamiltonian or on-shell spectrum  
 \ci{af,kmrz,review} (see also  \ci{call}  for earlier  hints of this symmetry at the level of the
 near plane wave spectrum)  but not as a manifest symmetry of the gauge-fixed 
 string action. 
}

A  possible  alternative to the existence of such an ``on-shell''
global fermionic target  space symmetry  is  a hidden 
  global 2-d   supersymmetry.  
Indeed, there is at least one special  case in which  the reduced-theory S-matrix
has  2-d supersymmetry -- the truncation to the $AdS_2 \times S^2$ case  when the reduced 
 Lagrangian 
\rf{susy2} or \rf{s2} is equivalent to that of the (2,2) supersymmetric sine-Gordon model. 
In this case the $SO(4)$ indices in \rf{theone}  take just one value, 
and thus the $SU(2)$ indices $a,\dot a, \alpha, \dot \alpha$ in \rf{trans},\rf{sut} 
   also take a single  value 
(so that the $\mbb{T}$-matrix coefficients $T^{CD}_{AB}$ in \rf{kek}
simplify to $K_1+K_2, K_3+K_4$, etc).

In general,  the action \rf{act} or the $SU(2)^4$ invariant  quartic 
gauge-fixed Lagrangian \rf{sut}  cannot be directly 
invariant under  the standard 2-d  supersymmetry  since the bosons and the fermions are
 in different
representations of the the global symmetry group \ci{gt1,gt2,rtfin}. 
This objection does not, however, apply 
to the special $SO(4)$ invariant form of the gauge-fixed Lagrangian in \rf{theone}.
It would be very interesting to check   if the obvious linear 
2-d supersymmetry of the 
quadratic part of \rf{theone} (the same as in the plane-wave limit of the 
 \adss superstring action \ci{mett})
  extends also to the quartic interaction  level.\foot{By analogy with \rf{s2} which
   is a direct truncation of  \rf{theone}  we may again  expect
to find  (2,2) supersymmetry, cf. Appendix B in \ci{hit}.}  It should be 
straightforward to check for 2-d supersymmetry \ci{swi} 
 in the $SO(4)$ invariant form 
of the S-matrix given in Appendix C.

 As for  the  manifestly $SU(2)^4$ invariant form of the action \rf{sut}  where 
the basic fields  have indices of  bi-fundamental representations  of pairs 
of different $SU(2)$ groups,  one  could   hope to realise a  2-d supersymmetry if one could 
represent them as products  of  4+4   ``pre-fields''  each  transforming in the fundamental representation
of one  of the four  $SU(2)$ groups, i.e., symbolically,\foot{A somewhat 
similar bi-linear representation
 was mentioned in footnote \ref{foot2}, which commented on  group  symmetry factorization of the S-matrix (cf.   \ci{kmrz,afz}).}   
\ $Y_{a\adot} = B^i_a  C^i_{\adot}, \ 
Z_{\al \ald} = F^i_\al  G^i_{\ald}$,   and similarly  for 
$\zel{}_{a\ald},\zer{}_{a\ald},$  
$\ch{}_{\al\adot},\chr{}_{\al\adot}. $  
Here $B,C,F,G$ are fermions and $\zeta$ and $\chi$ should  be built of one fermion and one boson 
 Then a  2-d supersymmetry may  be relating these  doublet ``pre-fields'' having the same type 
 of $SU(2)$ index. 
 It may happen that  some bosonisation/fermionization 
 transformation applied to  
 \rf{sut} may produce  a  local  Lagrangian for these ``pre-fields'' 
 which would be  invariant under  such  2-d supersymmetry.

As was mentioned  in the  Introduction, 
in attempting to find   an   exact quantum solution of the  reduced theory \rf{act} 
it is  natural to try to draw lessons from   examples  of massive integrable  deformations of coset CFT's 
already studied  in the literature  \ci{csd,mh,hsg} (see \ci{coaphd} for a review). 
Almost all of the previous papers on this  subject followed  \ci{hsg} in considering  $G/H$ models
 with an abelian  subgroup  $H$. 
The massive  deformations of the $G/H$   gWZW  models investigated  in \ci{hsg} 
were selected so that to have two properties: 
  (i)  small  fluctuations near  a vacuum should have massive spectrum, 
and (ii)  all  possible flat  directions of the potential 
should  correspond only to gauge transformations. 
Since $H$ is the global symmetry of the action, any constant matrix  
\be\la{vuc}
g_{vac}=h_0\in H\,,\hs{30pt}h_0=\,\trm{const}\,,
\ee 
 represents a minimum 
of the potential in \rf{bact}. Then the  condition  (ii)  is equivalent  to the 
requirement   that the vacuum configuration is unique up to a gauge transformation and 
this then restricts $H$ to  be  {\it abelian}   and in addition one is to use an asymmetric
(axial)  gauging of the WZW  model \ci{hsg}.    Such   
(``symmetric space sine-Gordon'')  models then have the complex sine-Gordon model
as its special case and have  features similar to those of the latter  theory  \ci{csd}: 
  they have  no vacuum degeneracy (no spontaneous symmetry breaking) 
and  thus may have  non-topological solitons only.
The latter  carry  abelian charges  and in the small  charge limit smoothly reduce 
to the elementary field excitations  around the unique vacuum. 
This implies,  in particular,   that 
their   exact   S-matrix  should  have a smooth limit 
 in which  it  reduces to the perturbative   S-matrix for the 
 elementary excitations (cf. \ci{csd}). 

 For  a non-abelian $H$  as  in case of  the \adss reduced  model \rf{co},\rf{act}
 discussed  in this paper there is
     an $H$-orbit of  a priori inequivalent vacua \rf{vuc}
   and thus  the corresponding 
    theory  is expected to have topological
    solitons which interpolate between these different vacua. As their  quantization 
    in 2-d is   a     potentially   intricate  problem,  it is not  clear if 
 a  Lagrangian description is enough in this case 
to construct the corresponding exact quantum  S-matrices.\foot{
We are grateful to  J.L. Miramontes  for related comments and 
important explanations of  these issues.}

An important feature  of the present model  \rf{act} that may help to  by-pass this 
potential  complication  is that it should  be viewed  not  just on its own but as a tool 
 for  solving    the  original \adss superstring theory. 
 The requirement of equivalence to the superstring theory 
  may  provide an   extra  input to define the  quantum version of the reduced theory. 
  Indeed,  as discussed in  \ci{gt1,hit}, 
  all of these  vacua of the reduced theory 
  correspond to the same  string configuration -- the ``plane-wave'' or   BMN vacuum of  
the original  superstring theory.\foot{This suggests, in particular,   
that  apparent  topological solitons of  the reduced theory  should either  be unphysical 
 or, more generally,   translate into non-topological solitons in the  superstring  theory
 (cf. relation between the sine-Gordon soliton   of reduced theory and the  giant magnon 
 soliton of string theory  \ci{hm}).}
Recall  \ci{gt1,hit} that when carrying out the reduction 
procedure  by starting with the first-order form of classical superstring equations 
one has  initially   $H\x H$ gauge symmetry $g\to h^{-1} g h', $ etc., 
  with independent $h,h' \in H$. Half of that ``on-shell'' gauge symmetry is then gauge-fixed 
  to get the reduced equations in the form that can be derived from a local
   action \rf{act}.\foot{This fixing is not unique, and different gauge fixings give rise 
   (to   apparently equivalent) reduced theories  based on  asymmetric gauging 
   of the remaining group $H$. In the non-abelian case there is a class of asymmetric gaugings (for which the $\tau$-automorphism
takes the form $\tau(u) = h_0^{-1} u h_0 $) that
   corresponds to effectively  changing  one vacuum for  another \ci{hit}:
   starting with  the symmetrically gauged WZW action   with $\tau= \bf 1$ 
and expanding it near $g= h_0$ is  equivalent to starting with  asymmetrically gauged action 
  containing  \rf{asymact}  with   $\tau(u) = h_0^{-1} u h_0 $ 
and expanding it near $g= \bf 1$.
    Such  different gaugings were  discussed already  in \cite{hsg} and 
their relation (in the case of abelian $H$) to T-duality 
was further clarified in \cite{mira1}.}
 Before this  partial gauge fixing it is always possible to choose the classical vacuum 
 to be the identity, $g=\bf 1$, i.e. all choices in \rf{vuc} are gauge-equivalent. 
 While this  is  no longer so at the level of the reduced theory action \rf{act} 
 one may argue that the information 
 in the reduced theory which is relevant for the original superstring 
 should not be sensitive to a particular choice of $h_0$ in \rf{vuc}.  
In this case  the reduced theory S-matrix   computed in this paper 
by expanding  near $g=\id$  should thus have a universal  meaning.

One of the central issues 
is that of a precise relation between the 
string-theory  and the reduced-theory   observables and parameters. 
In particular, in 
 comparing the reduced  and the superstring
  S-matrices in section 4  we assumed, following \ci{rtfin}, 
  that the 
overall constant $k$  in \rf{act} is directly related to the  
string tension $\sqrt \lambda \over 2 \pi$. 
However, in a  G/H  gauged WZW model with  compact $G$  or at least  compact non-abelian $H$ 
 the constant $k$  should  be quantized \ci{wite};  that quantization plays an important role 
 also in the present context of massive deformations of gWZW  models 
 as  was  emphasized in \ci{csd,mh}. 
At the same time, there is  no reason to expect quantization of the string tension
in the original \adss  string theory.  
One possible way out 
 is to assume that $k$ is an  additional hidden parameter 
while the string tension should enter  only through $\mu$, i.e.  
 in a combination with an arbitrary  mass scale.
Given that the  classical  target-space symmetry  charges (or rather Casimirs) 
of the original string theory  are ``hidden'' in the
 reduced theory,  the precise translation between  the observables should be   non-trivial. 
 Then  the string  tension may not appear  directly in the reduced theory action  but 
 rather in  the corresponding 
   definition of string observables in terms of reduced theory ones.

\section*{Acknowledgments}
We are grateful to  G. Arutyunov, N. Beisert,  Y. Iwashita, T. McLoughlin,   J.L. Miramontes,
 F. Spill and   A. Rej   for helpful  discussions.  
AAT would like to acknowledge   A. Alexandrov and especially 
R. Roiban   for   collaborations on related  issues. He also thanks R.Roiban 
for  many useful  comments
and suggestions. 
BH acknowledges the support of  EPSRC through his  studentship.


\appendices

\renewcommand{\theequation}{A.\arabic{equation}}

\setcounter{equation}{0}

\section{Simplification of quartic  Lagrangians
\la{A}}

Here we will  show how  to transform  the Lagrangian  
\eqref{l1} to \eqref{l2} using  integration by parts. 
We will also illustrate the steps (integration by parts and use of linearised equations of motion in the quartic terms) 
that allow one 
to put  \eqref{fulllagp} into the form 
\eqref{fulllag} leading to the equivalent 
 tree-level two-particle S-matrix. 
 Below  we will  ignore total derivatives and for 
 simplicity  of presentation omit the overall trace  (that 
 allows to rearrange terms using cyclicity property). 
 
The quartic term  in \eqref{l1} may be written as 
\be \bs 
&\fr{1}{8}\com{\M}{\dpl \M}\frac{\dm}{\dpl}\com{\M}{\dpl \M}+\fr{1}{4}\com{\dm \M}{\dpl \M}\fr{1}{\dpl}\com{\M}{\dpl \M}-\fr{1}{24}\com{\M}{\dpl \M}\com{\M}{\dm \M}
\nn \\&\quad=\fr{1}{8}\com{\M}{\dpl \M}\frac{\dm}{\dpl}\com{\M}{\dpl \M}+\fr{1}{4}\dm\left(\com{ \M}{\dpl \M}\right)\fr{1}{\dpl}\com{\M}{\dpl \M}
\la{df}
 \\& \quad\hs{20pt}-\fr{1}{4}\com{ \M}{\dpl \dm \M}\fr{1}{\dpl}\com{\M}{\dpl \M}-\fr{1}{24}\com{\M}{\dpl \M}\com{\M}{\dm \M}
\nn \\&\quad=-\fr{1}{8}\com{\M}{\dpl \M}\frac{\dm}{\dpl}\com{\M}{\dpl \M}-\fr{1}{4}\com{ \M}{\dpl \dm \M}\fr{1}{\dpl}\com{\M}{\dpl \M}-\fr{1}{24}\com{\M}{\dpl \M}\com{\M}{\dm \M}\,.
\nn
\end{split}\ee
Using that  
\be\la{mident}\bs
\frac{\dm}{\dpl}\com{\M}{\dpl \M}&=\fr{1}{\dpl}\big(\com{\dm \M}{\dpl \M}+\com{\M}{\dpl \dm \M}\big)
\\&=\fr{1}{\dpl}\big(\dpl\com{\dm \M}{\M}-2\com{\dpl \dm \M}{\M}\big)=-\com{\M}{\dm \M}+2\fr{1}{\dpl}
\com{\M}{\dpl\dm \M}\,,
\end{split}\ee
we find that  the non-local parts in the quartic term cancel and we end up 
with the  single term 
\be
\fr{1}{12}\com{\M}{\dpl \M}\com{\M}{\dm \M}\,.
\ee
Let us  now  discuss   the fermionic terms  in \eqref{fulllagp}.
Here we will use linearised equations of motion to simplify the quartic terms.
For example, since $\dpl \dm \M \propto \M$,  we may set terms with $[\M, \del_+ \del_- \M]$ to zero. 
Using  the linearised fermionic equations of motion \rf{eom}
and the fact that $T$ anticommutes with $\psr$ and $\psl$ we may show that 
\be\bs
\com{\psr}{\psl}&=\fr{4}{\mu^2}\com{T\dpl\psl}{T\dm\psr}=-\fr{4 T^2}{\mu^2}\com{\dpl\psl}{\dm\psr}
\\&=\fr{1}{\mu^2}\dpl\com{\psl}{\dm\psr}-\fr{1}{\mu^2}\com{\psl}{\dpl\dm \psr}=\fr{1}{\mu^2}\dpl\com{\psl}{\dm\psr}+\com{\psl}{\psr}\,.
\end{split}\ee
This 
 implies that we can  make the formal   substitutions 
in the   quartic terms in the action 
\be\la{psiident1}
\com{\psr}{\psl}\ \to \ \fr{1}{2\mu^2}\dpl\com{\psl}{\dm\psr}\,, \ \ \ \ \ \ \ \ 
\com{\psr}{\psl} \ \to \ \fr{1}{2\mu^2}\dm\com{\dpl\psl}{\psr}\,.
\ee
We can then make the following  transformations 
\be \bs
&\fr{1}{4}\com{\M}{\dpl \M}\fr{\dm}{\dpl}\com{\psr}{\psr T}+\fr{1}{4}\com{\psr}{\psr T}\fr{\dm}{\dpl}\com{\M}{\dpl \M}+\fr{1}{2}\com{\dm \M}{\dpl \M}\fr{1}{\dpl}\com{\psr}{\psr T}
\\&\quad=\fr{1}{2}\com{\M}{\dpl \M}\fr{\dm}{\dpl}\com{\psr}{\psr T}+\fr{1}{2}\dm\left(\com{\M}{\dpl \M}\right)\fr{1}{\dpl}\com{\psr}{\psr T}
\\&\quad\quad=\dm\Big(\fr{1}{2}\com{\M}{\dpl \M}\fr{1}{\dpl}\com{\psr}{\psr T}\Big)=0
\,,
\end{split}\ee
where we have dropped a term containing $\com{\M}{\dpl\dm \M}$, integrated by parts
and ignored all total derivatives.
To give another example  consider 
\be\la{hw}
\fr{\mu}{2}\com{\psr}{\psl}\fr{1}{\dpl}\com{\M}{\dpl \M}\,.
\ee
Using \eqref{psiident1},\eqref{mident} this term may  be written as
\be\bs
&\fr{1}{8\mu}\dpl\com{\psl}{\dm\psr}\fr{1}{\dpl}\com{\M}{\dpl \M}+\fr{1}{8\mu}\dm\com{\dpl\psl}{\psr}\fr{1}{\dpl}\com{\M}{\dpl \M}
\\=&-\fr{1}{8\mu}\com{\psl}{\dm\psr}\com{\M}{\dpl \M}+\fr{1}{8\mu}\com{\dpl\psl}{\psr}\com{\M}{\dm \M}\,,
\end{split}\ee
where integration by parts and the identity \rf{mident}  have also been used.
Taking into account  the linearised fermionic equations of motion
we finally get 
\be\bs
-\fr{1}{8}\com{2 \psl T}{\psl}\com{\M}{\dpl \M}-\fr{1}{8}\com{\psr}{2 T \psr}\com{\M}{\dm \M}\,.
\end{split}
\ee

\renewcommand{\theequation}{B.\arabic{equation}}
\setcounter{equation}{0}

\section{Basis for subspaces of $\psu$\la{B}}

Here we explicitly write out the basis for the relevant parts 
($\mathfrak{f}^\parallel_2$, $\mathfrak{f}^\parallel_1$ and $\mathfrak{f}^\parallel_3$)
of 
the superalgebra $\psu$ that we used in section \ref{index}.

An arbitrary element of bosonic subspace 
 $\mathfrak{f}^\parallel_2$  can be written as 
\begin{small}
\bea\nn &&f^\parallel_2\left(x_1,\,x_2,\,x_3,\,x_4,\,x_5,\,x_6,\,x_7,\,x_8\right)\,=\\&&\nn\;\;\left(
\ba{cccccccc}
0&0&x_1+i x_2&-x_3-i x_4&0&0&0&0
\\0&0&-x_3+i x_4&-x_1+i x_2&0&0&0&0
\\x_1-i x_2&-x_3-i x_4&0&0&0&0&0&0
\\-x_3+i x_4&-x_1-i x_2&0&0&0&0&0&0
\\0&0&0&0&0&0&x_5+i x_6&x_7+i x_8
\\0&0&0&0&0&0&-x_7+i x_8&x_5-i x_6
\\0&0&0&0&-x_5+i x_6&x_7+i x_8&0&0
\\0&0&0&0&-x_7+i x_8&-x_5-i x_6&0&0
\ea\right)\,,\eea
\end{small}
where $x_i$ are commuting parameters. An arbitrary element of fermionic subspace 
$\mathfrak{f}^\parallel_1$  is 
\begin{small}
\bea\nn &&f^\parallel_1\left(\alpha_1,\,\alpha_2,\,\alpha_3,\,\alpha_4,\,\alpha_5,\,\alpha_6,\,\alpha_7,\,\alpha_8\right)\,=\\&&\nn\;\;\left(
\ba{cccccccc}
0&0&0&0&0&0&\al_1+i\al_2&\al_3+i\al_4
\\0&0&0&0&0&0&-\al_3+i\al_4&\al_1-i\al_2
\\0&0&0&0&\al_5-i\al_6&-\al_7-i\al_8&0&0
\\0&0&0&0&-\al_7+i\al_8&-\al_5-i\al_6&0&0
\\0&0&-i\al_5+\al_6&i\al_7-\al_8&0&0&0&0
\\0&0&i\al_7+\al_8&i\al_5+\al_6&0&0&0&0
\\i\al_1+\al_2&-i\al_3+\al_4&0&0&0&0&0&0
\\i\al_3+\al_4&i\al_1-\al_2&0&0&0&0&0&0
\ea\right)\,,\eea
\end{small}
where $\al_i$ are anticommuting parameters.
Then an arbitrary element of fermionic $\mf{f}^\parallel_3$ is 
\bd
f^\parallel_3\left(\alpha_1,\,\alpha_2,\,\alpha_3,\,\alpha_4,\,\alpha_5,\,\alpha_6,\,\alpha_7,\,\alpha_8\right)=2 T f^\parallel_1\left(\alpha_1,\,\alpha_2,\,\alpha_3,\,\alpha_4,\,\alpha_5,\,\alpha_6,\,\alpha_7,\,\alpha_8\right)
\ed
Explicitly, we  choose  the following bases:

\nin for $\mathfrak{f}^\parallel_2$
\bd\ba{cc}
T^A_1=f^\parallel_2\left(\fr{1}{2},\,0,\,0,\,0,\,0,\,0,\,0,\,0\right)\,,\ \ \ &T^A_2=f^\parallel_2\left(0,\,\fr{1}{2},\,0,\,0,\,0,\,0,\,0,\,0\right)\,,
\\T^A_3=f^\parallel_2\left(0,\,0,\,\fr{1}{2},\,0,\,0,\,0,\,0,\,0\right)\,,\ \ \ &T^A_4=f^\parallel_2\left(0,\,0,\,0,\,\fr{1}{2},\,0,\,0,\,0,\,0\right)\,,
\\T^S_1=f^\parallel_2\left(0,\,0,\,0,\,0,\,\fr{1}{2},\,0,\,0,\,0\right)\,,\ \ \ &T^S_2=f^\parallel_2\left(0,\,0,\,0,\,0,\,0,\,\fr{1}{2},\,0,\,0\right)\,,
\\T^S_3=f^\parallel_2\left(0,\,0,\,0,\,0,\,0,\,0,\,\fr{1}{2},\,0\right)\,,\ \ \ &T^S_4=f^\parallel_2\left(0,\,0,\,0,\,0,\,0,\,0,\,0,\,\fr{1}{2}\right)\,,
\ea\ed
 for $\mf{f}^\parallel_1$
\bd\ba{cc}
T^{R_1}_1=f^\parallel_1\left(\fr{1}{2},\,0,\,0,\,0,\,0,\,0,\,0,\,0\right)\,,\ \ \ &T^{R_1}_2=f^\parallel_1\left(0,\,\fr{1}{2},\,0,\,0,\,0,\,0,\,0,\,0\right)\,,
\\T^{R_1}_3=f^\parallel_1\left(0,\,0,\,\fr{1}{2},\,0,\,0,\,0,\,0,\,0\right)\,,\ \ \ &T^{R_1}_4=f^\parallel_1\left(0,\,0,\,0,\,\fr{1}{2},\,0,\,0,\,0,\,0\right)\,,
\\T^{R_2}_1=f^\parallel_1\left(0,\,0,\,0,\,0,\,\fr{1}{2},\,0,\,0,\,0\right)\,,\ \ \ &T^{R_2}_2=f^\parallel_1\left(0,\,0,\,0,\,0,\,0,\,\fr{1}{2},\,0,\,0\right)\,,
\\T^{R_2}_3=f^\parallel_1\left(0,\,0,\,0,\,0,\,0,\,0,\,\fr{1}{2},\,0\right)\,,\ \ \ &T^{R_2}_4=f^\parallel_1\left(0,\,0,\,0,\,0,\,0,\,0,\,0,\,\fr{1}{2}\right)\,,
\ea\ed
for $\mf{f}^\parallel_3$
\bd \ba{cc}
T^{L_1}_1= f^\parallel_3\left(\fr{1}{2},\,0,\,0,\,0,\,0,\,0,\,0,\,0\right)\,,\ \ \ &T^{L_1}_2= f^\parallel_3\left(0,\,\fr{1}{2},\,0,\,0,\,0,\,0,\,0,\,0\right)\,,
\\T^{L_1}_3= f^\parallel_3\left(0,\,0,\,\fr{1}{2},\,0,\,0,\,0,\,0,\,0\right)\,,\ \ \ &T^{L_1}_4= f^\parallel_3\left(0,\,0,\,0,\,\fr{1}{2},\,0,\,0,\,0,\,0\right)\,,
\\T^{L_2}_1= f^\parallel_3\left(0,\,0,\,0,\,0,\,\fr{1}{2},\,0,\,0,\,0\right)\,,\ \ \ &T^{L_2}_2= f^\parallel_3\left(0,\,0,\,0,\,0,\,0,\,\fr{1}{2},\,0,\,0\right)\,,
\\T^{L_2}_3= f^\parallel_3\left(0,\,0,\,0,\,0,\,0,\,0,\,\fr{1}{2},\,0\right)\,,\ \ \ &T^{L_2}_4= f^\parallel_3\left(0,\,0,\,0,\,0,\,0,\,0,\,0,\,\fr{1}{2}\right)\,,
\ea\ed
These generators   satisfy a number of relations that we used 
 in section \ref{index} to write the Lagrangian  in  component form  \rf{theone}
\bd\ba{ccc}
\trm{STr}(T^A_m T^A_n)=\delta_{mn}\,,
&\trm{STr}(T^S_m T^S_n)=\delta_{mn}\,,
&\trm{STr}(T^A_m T^S_n)=0\,,
\ea\ed
\bd\ba{cccc}
\trm{STr}(T^{R_1}_m T^{L_1}_n)=\delta_{mn}\,,
&\trm{STr}(T^{R_2}_m T^{L_2}_n)=\delta_{mn}\,,
&\trm{STr}(T^{R_1}_m T^{L_2}_n)=0\,,
&\trm{STr}(T^{R_2}_m T^{L_1}_n)=0\,.
\ea\ed
\bd\ba{cc}
T^{L_1}_m=2T\ T^{R_1}_m\,,\ \ \ \ \ \ \ T^{R_1}_m=2 T^{L_1}_m\ T\,,\ \ \ 
\ \ \ \ \ T^{L_2}_m=2T\ T^{R_2}_m\,,\ \ \ \ \ \ T^{R_2}_m=2 T^{L_2}_m\ T\,,
\ea\ed 
\bd\ba{c}
\trm{STr}(\com{T^A_m}{T^A_n}\com{T^A_p}{T^A_q})=-\delta_{mp}\delta_{nq}+\delta_{mq}\delta_{np}\,,\ \ \ \ \ \ \ 
\trm{STr}(\com{T^S_m}{T^S_n}\com{T^S_p}{T^S_q})=\delta_{mp}\delta_{nq}-\delta_{mq}\delta_{np}
\ea\ed
\bd\ba{c}
\trm{STr}(\com{T^A_m}{\com{T^A_n}{T}}\com{T^A_p}{\com{T^A_q}{T}})=-\delta_{mn}\delta_{pq}\,,
\ \ \ \ \ \  \ \trm{STr}(\com{T^S_m}{\com{T^S_n}{T}}\com{T^S_p}{\com{T^S_q}{T}})=\delta_{mn}\delta_{pq}\,,
\ea\ed
\bd \ba{c}
\trm{STr}(\acom{T^{R_1}_m}{T^{L_1}_n}\com{T^A_p}{T^A_q})=-\fr{1}{2}\epsilon_{mnpq}-\fr{1}{2}\delta_{mp}\delta_{nq}+\fr{1}{2}\delta_{mq}\delta_{np}\,,
\\\trm{STr}(\acom{T^{R_1}_m}{T^{L_1}_n}\com{T^S_p}{T^S_q})=-\fr{1}{2}\epsilon_{mnpq}+\fr{1}{2}\delta_{mp}\delta_{nq}-\fr{1}{2}\delta_{mq}\delta_{np}\,,
\\\trm{STr}(\acom{T^{R_2}_m}{T^{L_2}_n}\com{T^A_p}{T^A_q})=\fr{1}{2}\epsilon_{mnpq}-\fr{1}{2}\delta_{mp}\delta_{nq}+\fr{1}{2}\delta_{mq}\delta_{np}\,,
\\\trm{STr}(\acom{T^{R_2}_m}{T^{L_2}_n}\com{T^S_p}{T^S_q})=\fr{1}{2}\epsilon_{mnpq}+\fr{1}{2}\delta_{mp}\delta_{nq}-\fr{1}{2}\delta_{mq}\delta_{np}\,,
\ea\ed
\bd \ba{c}
\trm{STr}(\com{T^A_m}{T^{R_1}_n}\com{T^A_p}{T^{L_1}_q})=\fr{1}{4}\left(\epsilon_{mnpq}-\delta_{mn}\delta_{pq}-\delta_{mp}\delta_{nq}+\delta_{mq}\delta_{np}\right)\,,
\\\trm{STr}(\com{T^S_m}{T^{R_1}_n}\com{T^S_p}{T^{L_1}_q})=\fr{1}{4}\left(\epsilon_{mnpq}+\delta_{mn}\delta_{pq}+\delta_{mp}\delta_{nq}-\delta_{mq}\delta_{np}\right)\,,
\\\trm{STr}(\com{T^A_m}{T^{R_2}_n}\com{T^A_p}{T^{L_2}_q})=\fr{1}{4}\left(-\epsilon_{mnpq}-\delta_{mn}\delta_{pq}-\delta_{mp}\delta_{nq}+\delta_{mq}\delta_{np}\right)\,,
\\\trm{STr}(\com{T^S_m}{T^{R_2}_n}\com{T^S_p}{T^{L_2}_q})=\fr{1}{4}\left(-\epsilon_{mnpq}+\delta_{mn}\delta_{pq}+\delta_{mp}\delta_{nq}-\delta_{mq}\delta_{np}\right)\,,
\\\trm{STr}(\com{T^A_m}{T^{R_1}_n}\com{T^S_p}{T^{L_2}_q})=\fr{1}{4}\left(-\epsilon_{mnpq}-\delta_{mn}\delta_{pq}+\delta_{mp}\delta_{nq}-\delta_{mq}\delta_{np}\right)\,,
\\\trm{STr}(\com{T^S_m}{T^{R_1}_n}\com{T^A_p}{T^{L_2}_q})=\fr{1}{4}\left(\epsilon_{mnpq}-\delta_{mn}\delta_{pq}+\delta_{mp}\delta_{nq}-\delta_{mq}\delta_{np}\right)\,,
\\\trm{STr}(\com{T^A_m}{T^{R_2}_n}\com{T^S_p}{T^{L_1}_q})=\fr{1}{4}\left(-\epsilon_{mnpq}+\delta_{mn}\delta_{pq}-\delta_{mp}\delta_{nq}+\delta_{mq}\delta_{np}\right)\,,
\\\trm{STr}(\com{T^S_m}{T^{R_2}_n}\com{T^A_p}{T^{L_1}_q})=\fr{1}{4}\left(\epsilon_{mnpq}+\delta_{mn}\delta_{pq}-\delta_{mp}\delta_{nq}+\delta_{mq}\delta_{np}\right)\,,
\ea\ed
\bd \ba{c}
\trm{STr}(\acom{T^{R_1}_m}{T^{L_1}_n}\acom{T^{R_1}_p}{T^{L_1}_q})=-\epsilon_{mnpq}\,,
\ \ \ \ \ \ \ \trm{STr}(\acom{T^{R_2}_m}{T^{L_2}_n}\acom{T^{R_2}_p}{T^{L_2}_q})=\epsilon_{mnpq}\,.
\ea\ed

\renewcommand{\theequation}{C.\arabic{equation}}
\setcounter{equation}{0}

\section{$\tm$-matrix of  reduced theory\\
  for $AdS_5 \times S^5$  superstring \la{C}}

Here we present the full expression for the  $\tm$-matrix corresponding to
the reduced Lagrangian \rf{theone} or \rf{sut} first in the $SO(4)$ notation
and then in the  manifest $SU(2)^4$ notation.  

\subsection*{$\tm$-matrix in $SO(4)$  form }

\tbf{Boson-Boson}
\begin{small}
\bea\nn
\mbb{T}\ket{Y^r(p_1)Y^s(p_2)}&=&\fr{1}{4\sinh\vt}\Big[-\left(\delta_m^r\delta_n^s+\cosh\vt\left(\delta_m^s\delta_n^r-\delta_{mn}\delta^{rs}\right)\right)\ket{Y^m(p_1)Y^n(p_2)}
\\\nn&&\hs{32pt}
+ \fr{1}{2}\sinh\fr{ \vt}{2}(\epsilon_{mn}^{\;\;\;\;\,rs}+\delta_m^r\delta_n^s-\delta_m^s\delta_n^r+\delta_{mn}\delta^{rs})\ket{\zeta^m(p_1)\zeta^n(p_2)}
\\\nn&&\hs{32pt}+ \fr{1}{2}\sinh\fr{ \vt}{2}(-\epsilon_{mn}^{\;\;\;\;\,rs}+\delta_m^r\delta_n^s-\delta_m^s\delta_n^r+\delta_{mn}\delta^{rs})\ket{\chi^m(p_1)\chi^n(p_2)}\Big]
\nn\\\nn
\mbb{T}\ket{Z^r(p_1)Z^s(p_2)}&=&\fr{1}{4\sinh\vt}\Big[\left(\delta_m^r\delta_n^s+\cosh\vt\left(\delta_m^s\delta_n^r-\delta_{mn}\delta^{rs}\right)\right)\ket{Y^m(p_1)Y^n(p_2)}
\\\nn&&\hs{32pt}
+ \fr{1}{2}\sinh\fr{ \vt}{2}(\epsilon_{mn}^{\;\;\;\;\,rs}-\delta_m^r\delta_n^s+\delta_m^s\delta_n^r-\delta_{mn}\delta^{rs})\ket{\zeta^m(p_1)\zeta^n(p_2)}
\\\nn&&\hs{32pt}+ \fr{1}{2}\sinh\fr{ \vt}{2}(-\epsilon_{mn}^{\;\;\;\;\,rs}-\delta_m^r\delta_n^s+\delta_m^s\delta_n^r-\delta_{mn}\delta^{rs})\ket{\chi^m(p_1)\chi^n(p_2)}\Big]
\nn\\\nn
\mbb{T}\ket{Y^r(p_1)Z^s(p_2)}&=&\fr{1}{4\sinh\vt}\Big[\fr{1}{2}\cosh \fr{\vt}{2}(-\epsilon_{mn}^{\;\;\;\;\,rs}-\delta_{mn}\delta^{rs}+\delta_m^r\delta_n^s+\delta_m^s\delta_n^r)\ket{\zeta^m(p_1)\chi^n(p_2)}
\\\nn&&\hs{32pt}+\fr{1}{2}\cosh \fr{\vt}{2} (-\epsilon_{mn}^{\;\;\;\;\,rs}+\delta_{mn}\delta^{rs}-\delta_m^r\delta_n^s-\delta_m^s\delta_n^r)\ket{\chi^m(p_1)\zeta^n(p_2)}\Big]
\eea
\\
\end{small}
\tbf{Fermion-Fermion}
\begin{small}
\bea\nn\mbb{T}\ket{\zeta^r(p_1)\zeta^s(p_2)}&=&\fr{1}{4\sinh\vt}\Big[\fr{1}{2}\sinh\fr{ \vt}{2}(\epsilon_{mn}^{\;\;\;\;\,rs}+\delta_m^r\delta_n^s-\delta_m^s\delta_n^r+\delta_{mn}\delta^{rs})\ket{Y^m(p_1)Y^n(p_2)}
\\\nn&&\hs{32pt}+\fr{1}{2}\sinh\fr{ \vt}{2}(\epsilon_{mn}^{\;\;\;\;\,rs}-\delta_m^r\delta_n^s+\delta_m^s\delta_n^r-\delta_{mn}\delta^{rs})\ket{Z^m(p_1)Z^n(p_2)}
\\\nn&&\hs{32pt}-\cosh\vt\;\epsilon_{mn}^{\;\;\;\;\,rs}\ket{\zeta^m(p_1)\zeta^n(p_2)}\Big]
\nn\\\nn
\mbb{T}\ket{\chi^r(p_1)\chi^s(p_2)}&=&\fr{1}{4\sinh\vt}\Big[\fr{1}{2}\sinh\fr{ \vt}{2}(-\epsilon_{mn}^{\;\;\;\;\,rs}+\delta_m^r\delta_n^s-\delta_m^s\delta_n^r+\delta_{mn}\delta^{rs})\ket{Y^m(p_1)Y^n(p_2)}
\\\nn&&\hs{32pt}+\fr{1}{2}\sinh\fr{ \vt}{2}(-\epsilon_{mn}^{\;\;\;\;\,rs}-\delta_m^r\delta_n^s+\delta_m^s\delta_n^r-\delta_{mn}\delta^{rs})\ket{Z^m(p_1)Z^n(p_2)}
\\\nn&&\hs{32pt}+\cosh\vt\;\epsilon_{mn}^{\;\;\;\;\,rs}\ket{\chi^m(p_1)\chi^n(p_2)}\Big]
\nn\\\nn
\mbb{T}\ket{\zeta^r(p_1)\chi^s(p_2)}&=&\fr{1}{4\sinh\vt}\Big[\fr{1}{2}\cosh \fr{\vt}{2}(-\epsilon_{mn}^{\;\;\;\;\,rs}-\delta_{mn}\delta^{rs}+\delta_m^r\delta_n^s+\delta_m^s\delta_n^r)\ket{Y^m(p_1)Z^n(p_2)}
\\\nn&&\hs{32pt}+\fr{1}{2}\cosh \fr{\vt}{2}(\epsilon_{mn}^{\;\;\;\;\,rs}-\delta_{mn}\delta^{rs}+\delta_m^r\delta_n^s+\delta_m^s\delta_n^r)\ket{Z^m(p_1)Y^n(p_2)}\Big]
\eea
\\
\end{small}
\tbf{Boson-Fermion}
\begin{small}
\bea\nn
\mbb{T}\ket{Y^r(p_1)\zeta^s(p_2)}
&=&\fr{1}{4\sinh\vt}\Big[\fr{1}{2}\big[\cosh\vt(-\epsilon_{mn}^{\;\;\;\;\,rs}-\delta_m^s\delta_n^r+\delta_{mn}\delta^{rs})
- \delta_{m}^r\delta_n^s\big]\ket{Y^m(p_1)\zeta^n(p_2)}
\\\nn&&\hs{32pt}+\fr{1}{2}\cosh \fr{\vt}{2}(\epsilon_{mn}^{\;\;\;\;\,rs}-\delta_m^r\delta_n^s+\delta_{mn}\delta^{rs}-\delta_m^s\delta_n^r)\ket{\zeta^m(p_1)Y^n(p_2)}
\\\nn&&\hs{32pt}+\fr{1}{2}\sinh\fr{\vt}{2}(-\epsilon_{mn}^{\;\;\;\;\,rs}-\delta_m^s\delta_n^r+\delta_{mn}\delta^{rs}+\delta_m^r\delta_n^s)
\ket{\chi^m(p_1)Z^n(p_2)}\Big]
\nn\\\nn
\mbb{T}\ket{Z^r(p_1)\zeta^s(p_2)}&=&\fr{1}{4\sinh\vt}\Big[\fr{1}{2}
\big[\cosh\vt(-\epsilon_{mn}^{\;\;\;\;\,rs}+\delta_m^s\delta_n^r-\delta_{mn}\delta^{rs})+\delta_{m}^r\delta_n^s\big]
\ket{Z^m(p_1)\zeta^n(p_2)}
\\\nn&&\hs{32pt}+\fr{1}{2}\cosh \fr{\vt}{2}(\epsilon_{mn}^{\;\;\;\;\,rs}+\delta_m^r\delta_n^s-\delta_{mn}\delta^{rs}+\delta_m^s\delta_n^r)\ket{\zeta^m(p_1)Z^n(p_2)}
\\\nn&&\hs{32pt}+\fr{1}{2}\sinh \fr{\vt}{2} (\epsilon_{mn}^{\;\;\;\;\,rs}-\delta_m^s\delta_n^r+\delta_{mn}\delta^{rs}+\delta_m^r\delta_n^s)\ket{\chi^m(p_1)Y^n(p_2)}\Big]
\nn\\\nn
\mbb{T}\ket{Y^r(p_1)\chi^s(p_2)}&=&\fr{1}{4\sinh\vt}\Big[\fr{1}{2}\big[\cosh\vt(\epsilon_{mn}^{\;\;\;\;\,rs}-\delta_m^s\delta_n^r+\delta_{mn}\delta^{rs})-\delta_{m}^r\delta_n^s\big]\ket{Y^m(p_1)\chi^n(p_2)}
\\\nn&&\hs{32pt}+\fr{1}{2}\cosh \fr{\vt}{2}(-\epsilon_{mn}^{\;\;\;\;\,rs}-\delta_m^r\delta_n^s+\delta_{mn}\delta^{rs}-\delta_m^s\delta_n^r)\ket{\chi^m(p_1)Y^n(p_2)}
\\\nn&&\hs{32pt}+\fr{1}{2}\sinh \fr{\vt}{2}(-\epsilon_{mn}^{\;\;\;\;\,rs}+\delta_m^s\delta_n^r-\delta_{mn}\delta^{rs}-\delta_m^r\delta_n^s)\ket{\zeta^m(p_1)Z^n(p_2)}\Big]
\nn\\\nn
\mbb{T}\ket{Z^r(p_1)\chi^s(p_2)}&=&\fr{1}{4\sinh\vt}\Big[\fr{1}{2}\big[\cosh\vt(\epsilon_{mn}^{\;\;\;\;\,rs}+\delta_m^s\delta_n^r-\delta_{mn}\delta^{rs})+\delta_{m}^r\delta_n^s\big]
\ket{Z^m(p_1)\chi^n(p_2)}
\\\nn&&\hs{32pt}+\fr{1}{2}\cosh \fr{\vt}{2}(-\epsilon_{mn}^{\;\;\;\;\,rs}+\delta_m^r\delta_n^s-\delta_{mn}\delta^{rs}+\delta_m^s\delta_n^r)\ket{\chi^m(p_1)Z^n(p_2)}
\\\nn&&\hs{32pt}+\fr{1}{2}\sinh \fr{\vt}{2}(\epsilon_{mn}^{\;\;\;\;\,rs}+\delta_m^s\delta_n^r-\delta_{mn}\delta^{rs}-\delta_m^r\delta_n^s)\ket{\zeta^m(p_1)Y^n(p_2)}\Big]
\eea
\end{small}

\subsection*{$\tm$-matrix  in $SU(2)^4$ form}

\tbf{Boson-Boson}
\begin{footnotesize}
\bea\nn
\mbb{T}\ket{Y_{a\adot}(p_1)Y_{b\bdot}(p_2)}&=&\fr{1}{4\sinh\vt}\Big[\left(2\sinh^2\fr{\vt}{2}\;\delta_a^c\delta_b^d\delta_{\adot}^{\cd}\delta_{\bdot}^{\dd}-\cosh\vt\left(\delta_a^c\delta_b^d\delta_{\adot}^{\dd}\delta_{\bdot}^{\cd}+\delta_a^d\delta_b^c\delta_{\adot}^{\cd}\delta_{\bdot}^{\dd}\right)\right)\ket{Y_{c\cd}(p_1)Y_{d\dd}(p_2)}
\\\nn&&\hs{32pt}-\sinh\fr{ \vt}{2}\left(\de_a^c\de_b^d\epsilon_{\adot\bdot}\epsilon^{\gad\ded}\ket{\zeta_{c\gad}(p_1)\zeta_{d\ded}(p_2)}+\epsilon_{ab}\epsilon^{\ga\de}\de_{\adot}^{\cd}\de_{\bdot}^{\dd}\ket{\chi_{\ga\cd}(p_1)\chi_{\de\dd}(p_2)}\right)\Big]
\\ \nn
\mbb{T}\ket{Z_{\al\ald}(p_1)Z_{\bet\betd}(p_2)}&=&\fr{1}{4\sinh\vt}\Big[\left(-2\sinh^2\fr{\vt}{2}\;\delta_{\al}^{\ga}\delta_{\bet}^{\de}\delta_{\ald}^{\gad}\delta_{\betd}^{\ded}+\cosh\vt\left(\delta_{\al}^{\ga}\delta_{\bet}^{\de}\delta_{\ald}^{\ded}\delta_{\betd}^{\gad}+\delta_\al^\de\delta_\bet^\ga\delta_{\ald}^{\gad}\delta_{\betd}^{\ded}\right)\right)\ket{Z_{\ga\gad}(p_1)Z_{\de\ded}(p_2)}
\\\nn&&\hs{32pt}+\sinh\fr{ \vt}{2}\left(\de_\al^\ga\de_\bet^\de\epsilon_{\ald\betd}\epsilon^{
\cd\dd}\ket{\chi_{\ga\cd}(p_1)\chi_{\de\dd}(p_2)}+\epsilon_{\al\bet}\epsilon^{cd}\de_{
\ald}^{\gad}\de_{\betd}^{\ded}\ket{\zeta_{c\gad}(p_1)\zeta_{d\ded}(p_2)}\right)\Big]
\\\nn
\mbb{T}\ket{Y_{a\adot}(p_1)Z_{\bet\betd}(p_2)}&=&\fr{1}{4\sinh\vt}\Big[\cosh\fr{\vt}{2}\left(\de_a^c\de_\bet^\de\de_{\adot}^{\dd}\de_{\betd}^{\gad}\ket{\zeta_{c\gad}(p_1)\chi_{\de\dd}(p_2)}-\de_a^d\de_\bet^\ga\de_{\adot}^{\cd}\de_{\betd}^{\ded}\ket{\chi_{\ga\cd}(p_1)\zeta_{d\ded}(p_2)}\right)\Big]
\eea
\\
\end{footnotesize}
\tbf{Fermion-Fermion}
\begin{footnotesize}
\bea\nn
\mbb{T}\ket{\zeta_{a\ald}(p_1)\zeta_{b\betd}(p_2)}&=&\fr{1}{4\sinh\vt}\Big[-\sinh\fr{ \vt}{2}\left(\de_a^c\de_b^d\epsilon_{\ald\betd}\epsilon^{\cd\dd} \ket{Y_{c\cd}(p_1)Y_{d\dd}(p_2)}-\epsilon_{ab}\epsilon^{\ga\de}\de_{\ald}^{\gad}\de_{\betd}^{\ded}\ket{Z_{\ga\gad}(p_1)Z_{\de\ded}(p_2)}\right) 
\\\nn&&\hs{32pt}+\cosh\vt\left(\de_a^c\de_b^d\de_{\ald}^{\ded}\de_{\betd}^{\gad}-\de_a^d\de_b^c\de_{\ald}^{\gad}\de_{\betd}^{\ded}\right)\ket{\zeta_{c\gad}(p_1)\zeta_{d\ded}(p_2)}\Big]
\\\nn
\mbb{T}\ket{\chi_{\al\adot}(p_1)\chi_{\bet\bdot}(p_2)}&=&\fr{1}{4\sinh\vt}\Big[-\sinh\fr{ \vt}{2}\left(\epsilon_{\al\bet}\epsilon^{cd}\de_{\adot}^{\cd}\de_{\bdot}^{\dd}\ket{Y_{c\cd}(p_1)Y_{d\dd}(p_2)}-\de_\al^\ga\de_\bet^\de\epsilon_{\adot\bdot}\epsilon^{\gad\ded}\ket{Z_{\ga\gad}(p_1)Z_{\de\ded}(p_2)}\right)
\\\nn&&\hs{32pt}+\cosh\vt\left(\de_{\al}^\de\de_\bet^\ga\de_{\adot}^{\cd}\de_{\bdot}^{\dd}-\de_{\al}^\ga\de_\bet^\de\de_{\adot}^{\dd}\de_{\bdot}^{\cd}\right)\ket{\chi_{\ga\cd}(p_1)\chi_{\de\dd}(p_2)}\Big]
\\\nn
\mbb{T}\ket{\zeta_{a\ald}(p_1)\chi_{\bet\bdot}(p_2)}&=&\fr{1}{4\sinh\vt}\Big[\cosh\fr{\vt}{2}\left(\de_a^c\de_\bet^\de\de_{\ald}^{\ded}\de_{\bdot}^{\cd}\ket{Y_{c\cd}(p_1)Z_{\de\ded}(p_2)}+\de_a^d\de_\bet^\ga\de_{\ald}^{\gad}\de_{\bdot}^{\dd}\ket{Z_{\ga\gad}(p_1)Y_{d\dd}(p_2)}\right)\Big]
\eea
\\
\end{footnotesize}
\tbf{Boson-Fermion}
\begin{footnotesize}
\bea\nn
\mbb{T}\ket{Y_{a\adot}(p_1)\zeta_{b\betd}(p_2)}&=&\fr{1}{4\sinh\vt}\Big[\Big(\sinh^2\fr{\vt}{2}\;\de_a^c\de_b^d\de_{\adot}^{\cd}\de_{\betd}^{\ded}-\cosh\vt\de_a^d\de_b^c\de_{\adot}^{\cd}\de_{\betd}^{\ded}\Big)\ket{Y_{c\cd}(p_1)\zeta_{d\ded}(p_2)}
\\\nn&&\hs{32pt}-\cosh\fr{\vt}{2}\;\de_a^c\de_b^d\de_{\adot}^{\dd}\de_{\betd}^{\gad}\ket{\zeta_{c\gad}(p_1)Y_{d\dd}(p_2)}-\sinh\fr{\vt}{2}\;\epsilon_{ab}\epsilon^{\ga\de}\de_{\adot}^{\cd}\de_{\betd}^{\ded}\ket{\chi_{\ga\cd}(p_1)Z_{\de\ded}(p_2)}\Big]
\\\nn
\mbb{T}\ket{Z_{\al\ald}(p_1)\zeta_{b\betd}(p_2)}&=&\fr{1}{4\sinh\vt}\Big[\Big(-\sinh^2\fr{\vt}{2}\;\de_\al^\ga\de_b^d\de_{\ald}^{\gad}\de_{\betd}^{\ded}+\cosh\vt\de_\al^\ga\de_b^d\de_{\ald}^{\ded}\de_{\betd}^{\gad}\Big)\ket{Z_{\ga\gad}(p_1)\zeta_{d\ded}(p_2)}
\\\nn&&\hs{32pt}+\cosh\fr{\vt}{2}\;\de_\al^\de\de_b^c\de_{\ald}^{\gad}\de_{\betd}^{\ded}\ket{\zeta_{c\gad}(p_1)Z_{\de\ded}(p_2)}-\sinh\fr{\vt}{2}\;\de_{\al}^\ga\de_b^d\epsilon_{\ald\betd}\epsilon^{\cd\dd}\ket{\chi_{\ga\cd}(p_1)Y_{d\dd}(p_2)}\Big]
\\\nn
\mbb{T}\ket{Y_{a\adot}(p_1)\chi_{\bet\bdot}(p_2)}&=&\fr{1}{4\sinh\vt}\Big[\Big(\sinh^2\fr{\vt}{2}\;\de_a^c\de_\bet^\de\de_{\adot}^{\cd}\de_{\bdot}^{\dd}-\cosh\vt\de_a^c\de_\bet^\de\de_{\adot}^{\dd}\de_{\bdot}^{\cd}\Big)\ket{Y_{c\cd}(p_1)\chi_{\de\dd}(p_2)}
\\\nn&&\hs{32pt}-\cosh\fr{\vt}{2}\;\de_a^d\de_\bet^\ga\de_{\adot}^{\cd}\de_{\bdot}^{\dd}\ket{\chi_{\ga\cd}(p_1)Y_{d\dd}(p_2)}+\sinh\fr{\vt}{2}\;\de_{a}^c\de_\bet^\de\epsilon_{\adot\bdot}\epsilon^{\gad\ded}\ket{\zeta_{c\gad}(p_1)Z_{\de\ded}(p_2)}\Big]
\\\nn
\mbb{T}\ket{Z_{\al\ald}(p_1)\chi_{\bet\bdot}(p_2)}&=&\fr{1}{4\sinh\vt}\Big[\Big(-\sinh^2\fr{\vt}{2}\;\de_\al^\ga\de_\bet^\de\de_{\ald}^{\gad}\de_{\bdot}^{\dd}+\cosh\vt\de_\al^\de\de_\bet^\ga\de_{\ald}^{\gad}\de_{\bdot}^{\dd}\Big)\ket{Z_{\ga\gad}(p_1)\chi_{\de\dd}(p_2)}
\\\nn&&\hs{32pt}+\cosh\fr{\vt}{2}\;\de_\al^\ga\de_\bet^\de\de_{\ald}^{\ded}\de_{\bdot}^{\cd}\ket{\chi_{\ga\cd}(p_1)Z_{\de\ded}(p_2)}+\sinh\fr{\vt}{2}\;\epsilon_{\al\bet}\epsilon^{cd}\de_{\ald}^{\gad}\de_{\bdot}^{\dd}\ket{\zeta_{c\gad}(p_1)Y_{d\dd}(p_2)}\Big]
\eea
\end{footnotesize}

\renewcommand{\theequation}{D.\arabic{equation}}
\setcounter{equation}{0}

\section{S-matrix of massive deformation 
of\\ geometric $G/H$ coset sigma model\la{D}}

It is of interest to compare the expanded Lagrangian  and the 
S-matrix of the massive theory based on the  $G/H$ gauged WZW  model \rf{bact} 
with those of a similar massive theory based on the standard $G/H$ coset model, i.e. 
\be \la{coc}
\mc{S}=- k\;\int d^2x \;  \Tr \Big[\  
\textstyle{\fr{1}{2}}\left(\gi\dpl g-A_+\right)\left(\gi\dm g-A_-\right)
+{\mu^2}\gi T gT  \Big]\,.
\ee
Here  we use the same notation  and definitions 
for $g,A_\pm$ and $T$ as in section 2.1. This action is invariant under 
the following gauge transformations
(cf. \rf{gag}) 
\be \la{gah}
g \to g h\ , \ \ \ \ \ \ 
A_\pm \to  h^{-1}  A_\pm  h + h^{-1} \del_\pm h \ , \ \ \ \ \ \ \ 
  h=h(x)  \in H \ . \ee
Compared to \rf{bact},   this theory,  however, is not integrable  for $\mu\not=0$
(the Lax pair that exists in the classical  massless coset theory    does not appear to have  a generalization
for   $\mu\not=0$). 
  
Going through the same steps as in section 2, i.e. expanding near $g=\id$ using 
\rf{lf1},\rf{decomp},  fixing the $A_+=0$ gauge and solving the $A_-$-constraint  for $\xi$
(which is again the same $(\gi \dpl g)_{\mf{h}}=0$ as in gWZW case)
 we  end up with the following counterpart of the quartic Lagrangian 
\rf{lmgwzw} (cf.  also \rf{l1},\rf{l2})
\be
\begin{split}
\Lag=&\fr{1}{2}\dpl \M_a \dm \M^a-\fr{\mu^2}{2}\M_a \M^a+\fr{1}{8k}\gamma_{abcd}\M^a \dpl \M^b \frac{\dm}{\dpl}\left(\M^c \dpl \M^d\right)
\\&-\fr{1}{24k}\gamma_{abcd}\M^a \M^c \dpl \M^b \dm \M^d+
\fr{ 1  }{24k } \g_0 \mu^2 
\M^a \M_a \M^b \M_b +  \mathcal{O} ( \M^5) \,.
\end{split}
\ee
Here $\g$'s are defined as in \rf{cou}. Using identities in Appendix \ref{A} it is again possible to put this Lagrangian into a local form
\be\la{acv}
\Lag=\fr{1}{2}\dpl \M_a \dm \M^a-\fr{\mu^2}{2}\M_a \M^a-\fr{1}{6k}\gamma_{abcd}\M^a \M^c \dpl \M^b \dm \M^d+
\fr{ 1  }{24k } \g_0 \mu^2 
\M^a \M_a \M^b \M_b +  \mathcal{O} ( \M^5) \,.
\ee
The  difference between  the two actions is due to the contribution of the WZ term in 
\rf{bact} (other terms in the two actions  are the same in the $A_+=0$ gauge). 
From \rf{acv} we get   the following $\mbb{T}$-matrix 
 defined in \rf{expansion},\rf{vetd} 
\be\label{smatcoset}\begin{split}
T_{ab}^{cd}\left(\vt\right)=&\fr{1}{12\sinh\vt}\Big[\g_0 \left(\delta_{ab}\delta^{cd}+\delta_{a}^{c}\delta_{b}^{d}+\delta_{a}^{d}\delta_{b}^{c}\right) -2\gamma_{ab}^{\;\;\;cd} +2\gamma_{a\;\,b}^{\;\,d\;\,c} \\&+ \ 2\left(\gamma_{ab}^{\;\;\;cd}
+2\gamma_{a\;\,b}^{\;\,c\;\,d}+\gamma_{a\;\,b}^{\;\,d\;\,c}\right)\cosh \vt\Big]\,,
\end{split}
\ee
which is  similar to the  gWZW  result  \rf{t}.

For example, if we  consider  $G/H=SO(N)/SO(N-1)=S^{N-1}$ 
embedded into $F=SO(N+1)$ then using the identities in section \ref{sons} \eqref{smatcoset} simplifies to
(here $a,b,c,d=1,\,\ldots,\, N-1$)
\be\label{smatc12}\begin{split}
T_{ab}^{cd}\left(\vt\right)=\fr{1}{4\sinh\vt}\Big[\left(1
+2\cosh \vt\right)\delta_{ab}\delta^{cd}-\delta_{a}^{c}\delta_{b}^{d}+\left(1
-2\cosh \vt\right)\delta_{a}^{d}\delta_{b}^{c}\Big]\,.
\end{split}
\ee
In the case where $N=3$ this   expression can be rederived by starting  with 
\rf{coc}  and     fixing the $H=SO(2)$  symmetry \rf{gah} 
by a  gauge condition  on $g$  (here $T_1,T_2$ are generators of the coset defined in sections \ref{setup} and \ref{sons})  
\be
g=e^{z_1 T_1+ z_2 T_2}\,, \ \ \ \ \ \ \ \ 
z_1 +i z_2 = \rho e^{i u}  \ . 
\ee
Then  integrating out $A_\pm$  gives 
\be
\Lag=\fr{1}{2}\dpl\rho \dm \rho +\sin^2\fr{\rho }{2}\, 
\dpl  u \dm  u +2\mu^2 \cos \fr{\rho }{\sqrt{2}}\,.
\ee
Expanding this Lagrangian 
to quartic order in $\vp$ or in ``cartesian'' coordinates $z_a=(z_1,z_2)$ 
we get  (cf. \rf{csg1},\rf{csge}) 
\be
\Lag=\fr{1}{2}\dpl z_a \dm z^a-\fr{\mu^2}{2}z_az^a -\fr{1}{12}
(\delta_{ac}\delta_{bd} - \delta_{ad}\delta_{bc}) 
 z^a z^c  \dpl z^b \dm z^d+\fr{\mu^2}{48}z_az^a z_b z^b+ \mathcal{O}(z^5)\,,
\ee
and this again leads to  \eqref{smatc12}. 

In the  $N=5$ case  the field $\M^a$ in \rf{acv} transforms in the vector
 representation of $H=SO(4)$. As this  vector representation  is equivalent
  to the bi-fundamental representation  of $SU(2)\x SU(2)$ we can 
  rewrite \rf{smatc12} in terms of the  $SU(2)$ 
  fundamental representation  indices $\alpha,\, \dot{\alpha}$  as in \rf{trans}
   Then \rf{smatcoset}    becomes 
\be\begin{split}
T_{\al\ald,\bet\betd}^{\ga\gad,\de\ded}\left(\vt\right)=&\fr{1}{4\sinh\vt}\Big[-(2\cosh\vt+1) \left(\delta_{\al}^{\ga}\delta_{\bet}^{\de}\delta_{\ald}^{\ded}\delta_{\betd}^{\gad}+\delta_{\al}^{\de}\delta_{\bet}^{\ga}\delta_{\ald}^{\gad}\delta_{\betd}^{\ded}\right)+2\cosh \vt\,\delta_{\al}^{\ga}\delta_{\bet}^{\de}\delta_{\ald}^{\gad}\delta_{\betd}^{\ded}
\\&\hs{200pt}+\ 2\delta_{\al}^{\de}\delta_{\bet}^{\ga}\delta_{\ald}^{\ded}\delta_{\betd}^{\gad}\Big]\,.
\end{split}
\ee
Comparing to \rf{tef} we see that due  the  term in the second line here 
 we do not have the group factorisation of the $\mbb{T}$-matrix that we had in the gWZW case.
 This is a reflection of the fact that the theory \rf{coc}  is not integrable.

 \


\begin{small}

  \end{small}
\end{document}